\documentclass[11pt]{article}

\usepackage{amsfonts,amsmath,amssymb,amsthm,url,verbatim,color,bm,algorithmic,subfig,bbm,enumerate}
\usepackage[boxed]{algorithm}
\usepackage{natbib}
 \bibpunct[, ]{[}{]}{,}{a}{}{,}%

\usepackage[top=0.9in,bottom=1.1in,left=2.05in,right=0.05in]{geometry}

\usepackage{ifpdf}
\ifpdf
\usepackage[pdftex]{graphicx}
\else
\usepackage[dvips]{graphicx}
\fi

\newcommand\E{\ensuremath{\mathbb{E}}}

\newcommand\cX{\ensuremath{\mathcal{Z}}}

\newcommand\cR{\ensuremath{\mathcal{X}}}

\newcommand\bam{\ensuremath{\mathcal{I}}}
\newcommand\bq{\ensuremath{\mathcal{Q}}}
\newcommand\balph{\ensuremath{\bar{\alpha}}}
\newcommand\vs{\ensuremath{\mathbf{s}}}

\DeclareMathOperator*{\diag}{diag}

\newcommand\imm{\ensuremath{{\iota}}}

\newcommand\R{\ensuremath{\mathbb{R}}}

\newcommand\F{\ensuremath{\mathcal{F}}}

\newcommand\PP{\ensuremath{\mathbb{P}}}

\newcommand\vp{ {\vec{\pi}}}

\newtheorem{prop}{Proposition}
\newtheorem{cor}{Corollary}

\title{Sequential Bayesian Inference in Hidden Markov Stochastic Kinetic Models with Application to Detection and Response to Seasonal Epidemics}
\medskip\medskip
\author{Junjing Lin and Michael Ludkovski \\ \small{Department of Statistics \& Applied Probability, UC Santa Barbara} \thanks{E-mail: {linj@pstat.ucsb.edu}, {ludkovski@pstat.ucsb.edu}}}
\date{\today}

\begin{document}
\maketitle

\begin{abstract}
We study sequential Bayesian inference in continuous-time stochastic kinetic models with latent factors. Assuming continuous observation of all the reactions, our focus is on joint inference of the unknown reaction rates and the dynamic latent states, modeled as a hidden Markov factor. Using insights from nonlinear filtering of jump Markov processes we develop a novel sequential Monte Carlo algorithm for this purpose. Our approach applies the ideas of particle learning to minimize particle degeneracy and exploit the analytical jump Markov structure. A motivating application of our methods is modeling of seasonal infectious disease outbreaks represented through a compartmental epidemic model. We demonstrate inference in such models with several numerical illustrations and also discuss predictive analysis of epidemic countermeasures using sequential Bayes estimates.
\end{abstract}

{Keywords: sequential Monte Carlo, particle learning, jump Markov process, stochastic epidemic models}

\section{Introduction}\label{sec:intro}
Stochastic jump-Markov models have become ubiquitous in multiple application areas, including systems biology, molecular chemistry, epidemiology, queuing theory and finance. An important class of such systems is described using the chemical reaction system paradigm, which classifies jumps in terms of a finite number of possible reactions. System transitions are specified probabilistically in terms of the distributions of the inter-reaction periods and next-to-fire reaction type. The reaction rates depend on the current system state and impose a Markovian structure, termed a stochastic kinetic model (SKM) \citep{Wilk:stoc:2006}.

While the basic setup assumes time-stationarity, in many contexts time-dependence, seasonality, and other regime shifts of reaction rates are crucial. A popular way to incorporate {stochastic} shifts in the environment is through Markov modulation, i.e.~introduction of an additional (latent) dynamic factor that affects transition rates. We christen such systems \emph{Hidden Markov Stochastic Kinetic Models} (HMSKM).

Usage of an HMSKM in an application requires statistical estimation of the reaction rates and environmental factors. In the present paper we are concerned with Bayesian inference which allows unified filtering of latent system states and parameters and full quantification of the posterior uncertainty. Moreover, anticipating dynamic optimization and decision making applications, we are interested in sequential inference. As a motivating example of such HMSKM inference, we describe below a stochastic model of \emph{seasonal epidemics} of infectious diseases. Here the SKM paradigm is used to give a mechanistic description of outbreak progression using a compartmental description of the population, while the latent factor represents environmental factors affecting the outbreak (such as new genetic shifts in the pathogen or weather patterns). Sequential inference of the reaction rates (infectiousness, etc.) and dynamic seasonality determines outbreak severity and is the central problem in biosurveillance and corresponding public health policy-making.

Bayesian inference in SKMs is typically accomplished through Markov chain Monte Carlo (MCMC) methods \citep{Goli:Wilk:baye:2006,Boys:Wilk:Kirk:baye:2008,NiemiThesis,GolightlyWilkinson11} and has been also used for related epidemics inference problems  \citep{ONeill02,JewellKypraiosNealRoberts09, lawson2009bayesian,MerlGramacy09}. But sequential inference requires re-running an MCMC analysis after every new data point is obtained and therefore is not computationally efficient.
A more suitable and flexible alternative is to apply sequential Monte Carlo methods (SMC), also known as particle filters, which use an empirical selection-mutation mechanism \citep{deFreitas,RydenBook}. The main drawback of SMC is particle degeneracy which becomes especially serious during estimation of constant parameters. A recent class of \emph{particle learning} methods \citep{CarvalhoPL11,DukicPolsonPL} has been designed to overcome these challenges by exploiting additional analytical structures. Starting with these ideas we develop new SMC methods targeting continuous-time stochastic kinetic models, demonstrating the efficiency and tractability of Bayesian inference in this novel context.

The aims of this work are three-fold. First, we wish to draw the attention of the computational statistics and stochastic simulation community to the convenient analytic structure of hidden Markov stochastic kinetic models. This structure allows for highly efficient Bayesian sequential inference algorithms in rather general settings, which to our knowledge supercede previous results in this direction.
Second, we provide a new extension of the {particle learning} SMC method. In contrast to existing literature, we work in continuous time and further consider a hidden Markov factor. Thus, to compute the predictive and conditional likelihoods we use tools from stochastic filtering (namely filtering of doubly-stochastic Poisson processes) which have not been hitherto used in this context. Our results demonstrate the wide applicability and attractiveness of particle learning in jump-Markov models.

Finally, we extend the burgeoning body of literature on sequential inference in compartmental epidemiological models. Precisely, we propose a new SIR-type model of seasonal epidemics with a stochastic seasonality factor. The model allows for tractable online inference assuming full observation of all epidemiological events and can be used as a testbed to analyze alternative ways of epidemic control under imperfect information.

In the next section we provide more background on our main motivating application in epidemic modeling;
the rest of the paper is organized as follows. Section \ref{sec:general} provides a general setup of a hidden Markov stochastic kinetic model; the resulting inference problem and the particle learning algorithm are constructed in Section \ref{sec:smc}. Section \ref{sec:sis} then illustrates the results and provides numerical examples for a simple SIS model of seasonal epidemics. Section \ref{sec:policy}  then discusses predictive analysis of outbreak countermeasures in the developed sequential framework.

\subsection{Seasonal Epidemics}
Probabilistic modeling of infectious disease epidemics is an important tool in public health analysis. Stochastic models provide a tractable way to quantify the uncertainty about epidemic dynamics and carry out predictive analysis on the future path of the outbreak. Public health agencies increasingly rely on such techniques to compare alternative courses of action and mount an efficient policy response through behavioral or biomedical interventions.

\begin{figure}
  \begin{center}
   \includegraphics[width=0.59\textwidth,trim=0.05in 0.25in 0in 0.1in]{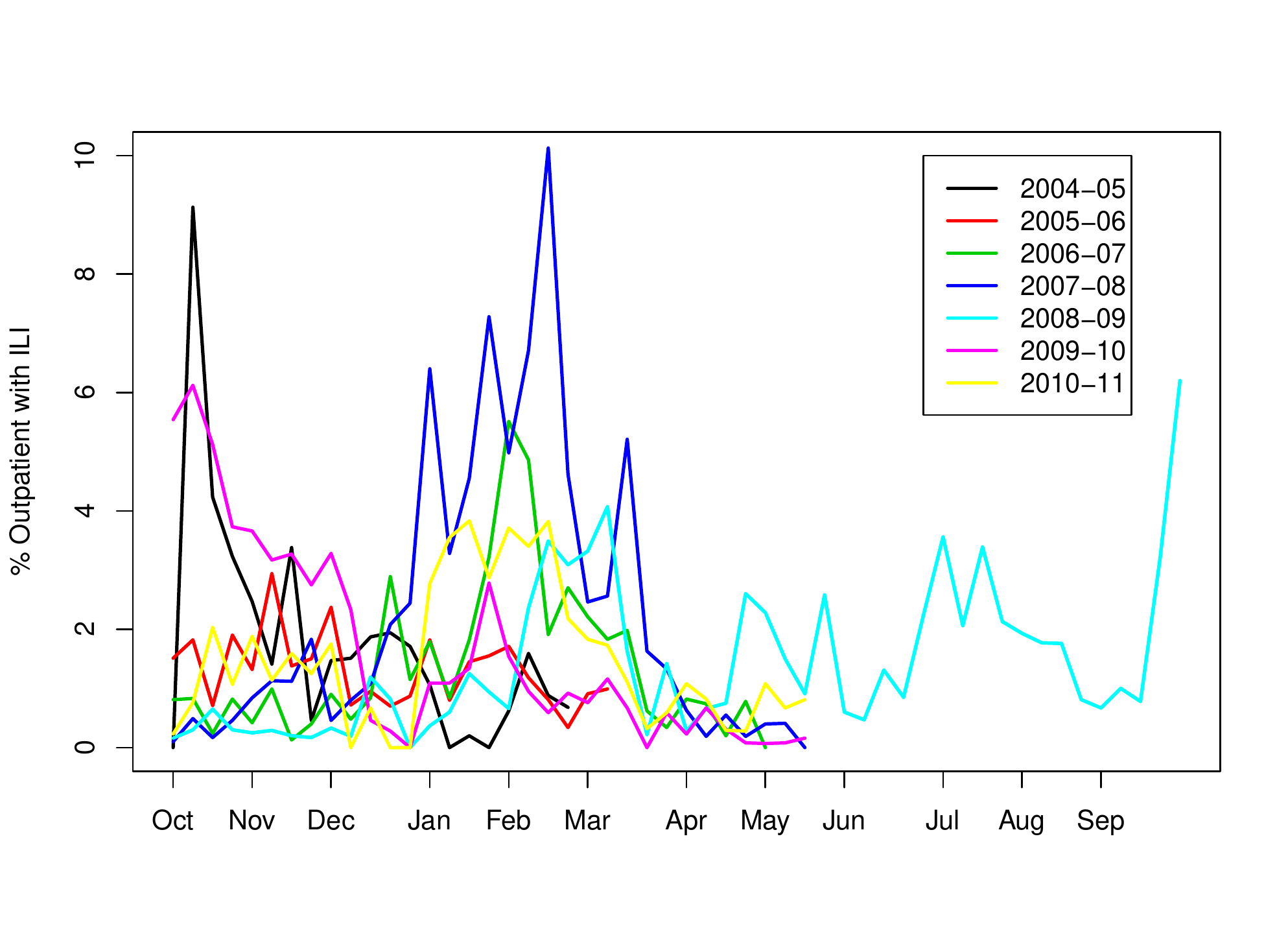}
  \caption{Percentage of Influenza Like Illness (ILI) outpatient visits in Santa Barbara county in 2004-2011 based on weekly sentinel providers data collected by Santa Barbara County Department of Public Health \citep{Sbc-Phd-report}. No data is normally collected during the summer. \label{fig:seasonal-flu}}
  \end{center}
\end{figure}

Recurring seasonal epidemics, with a prime example of influenza, provide some of the best testbeds for quantitative analysis thanks to availability of time-series data of previous outbreaks. Figure \ref{fig:seasonal-flu} shows the ILI flu incidence statistic in Santa Barbara county in California over the past seven years. ILI or influenza-like illness is a formal description of symptoms that typically occur in infected individuals, and is the most common proxy for measuring the incidence of flu in the population.
The figure illustrates the apparent fact that flu is generally more prevalent during winters, due to emergence of new virus strains and colder weather making susceptibility higher.
Nevertheless, significant variation in flu incidence cases and timing of outbreaks can be observed year over year. A case in point was the 2008-09 season when the new H1N1 influenza strain caused a world-wide pandemic peaking in early Fall 2009, scrambling the traditional flu calendar of public health actions\footnote{The usual H3N1 strain was also present that year and hence the time series effectively combines two distinct outbreaks}.  To summarize, Figure \ref{fig:seasonal-flu} demonstrates three main features of flu outbreaks: (i) strong seasonality; (ii) year-to-year variability; and (iii) stochastic fluctuations during each outbreak. Other endemic infectious diseases with similar patterns include rotavirus, norovirus, measles, dengue fever, and cholera \citep{GrasslyFraser06}.

Full understanding of community epidemic dynamics remains elusive given the paucity of available data. While issues such as inhomogeneous mixing, age effects, and spatial interactions are undoubtedly crucial, their mathematical and statistical modeling requires large-scale computational and modeling efforts \citep{HalloranPnas08}. Alternatively, \emph{mechanistic} models of outbreak provide a simplified but highly tractable paradigm of describing outbreak progression that can be calibrated to real data.  A popular mechanistic approach is given by the class of stochastic compartmental models \citep{AnderssonBrittonBook}. Thus, the population is partitioned into several classes of individuals based on their epidemiological status, such as Susceptible, Infected, etc., and the outbreak is described on the macroscopic level in terms of \emph{transition rates} $\Theta$ among the compartments. This SIR framework has been successfully used in modeling a range of infectious diseases, ranging from influenza \citep{MerlGramacy09} to measles \citep{CauchemezFerguson08}, and foot-and-mouth disease \citep{JewellKypraiosNealRoberts09}. Probabilistically, this approach corresponds to imposing a Markovian structure at the group level and captures the intrinsic uncertainty through the stochasticity of the transitions taking place.

To model the observed strong seasonality of influenza outbreaks, we introduce a further level of uncertainty through a stochastic seasonal factor $\{ M_t\}$. Thus, the transition rates between the compartmental classes are modulated by $\{ M_t\}$, which can be interpreted as the seasonal presence of a new pathogen. This seasonal factor is evidently an abstract object, i.e.~latent\footnote{As stated by \cite{GrasslyFraser06} ``despite the near ubiquity of this phenomenon [seasonality], the causes and consequences of seasonal patterns of incidence are poorly understood''}.
We note that its effect is indirect, since $\{M_t\}$ only affects the \emph{rates} of the transitions, not transitions themselves. We refer to \cite{LeStratCarrat,MartinezBeneito08} for related hidden Markov model representations of epidemics. There is further large literature on seasonal forcing in epidemics that focuses on deterministic SIR models using tools of dynamical systems \citep{KeelingRohaniGrenfell01,DushoffPlotkin04,StoneOlinky07} or multi-scale analysis \citep{KuskeGordilloGreenwood07}.

Bayesian inference in compartmental models consists of estimating the transition rates between classes and the dynamic size of each class.
This is by now a classical problem in biosurveillance; for example, a whole strand of literature is devoted
to estimating the basic reproductive ratio $\mathcal{R}$ which is the single most important parameter for predicting epidemic impact \citep{BallNeal02,CintronArias09,ONeill02,Chowell09adaptive}. The complementary inference of latent states is treated in the aforementioned \cite{LeStratCarrat,MartinezBeneito08}. However, little research has been done for joint parameter and state estimation
due to the associated computational challenges inherent in both Markov chain and sequential Monte Carlo  approaches. In this paper we present a novel algorithm for such joint inference of $\{M_t\}$ and outbreak parameters $\Theta$ using sequential Monte Carlo.

Beyond pure inference, the ultimate objective of  decision-makers is to mitigate the epidemic impact. This is achieved by implementing response policies including vaccination, quarantine, pharmaceutical or hospital treatment, information campaigns, etc. Since any policy involves budgetary or human resources, a balance is needed between costs due to epidemic morbidity and mortality and costs arising from policy actions. Moreover, to be optimal, a policy must be adaptive, i.e. rely on the latest collected information; thus, policy analysis is inherently linked to sequential inference. Quantitative approaches to such dynamic epidemic management include continuous-time Markov chain models \citep{MerlGramacy09}, Markov decision processes \citep{TannerSattenspielNtaimo08}, systems of ordinary differential equations \citep{Chowell09adaptive}, agent-based representations \citep{HalloranPnas08} and stochastic control \citep{LN10}.
 However, existing methods are limited in adequately addressing the questions of unknown system dynamics, parameters and states. 
Below we demonstrate that regime-switching compartmental models in fact provide a flexible paradigm for analyzing these issues by allowing for accurate sequential inference.

\section{General Setup}\label{sec:general}
 We consider a $d$-dimensional continuous-time jump-Markov process $\{ X_t\}$, $X_t \in \mathbb{N}^d$ on a probability space $(\Omega, \F, \PP)$, that evolves according to laws of a chemical reaction system. Namely, $X^k_t$ denotes the (non-negative) number of species of class $k=1,\ldots,d$ and there are $\bq$ reaction \emph{types} with corresponding stoichiometry vectors $\Delta_q \in \mathbb{Z}^d$, $q=1,2,\ldots, \bq$. The $\Delta_q$'s indicate the impact of a reaction on $X_t$: denoting by $\tau_k$, $k=1,\ldots,$ the reaction times, i.e.~the jump times of $\{X_t\}$, if the $k$-th reaction is of type $q$, then $X_{\tau_k} = X_{\tau_k-} + \Delta_q$. Between reactions $\{X_t\}$ is constant.

A convenient representation of $\{X_t\}$ is via a multivariate marked point process $\cR \equiv (\tau_k, R_k)$,
$k=1,2,\ldots,$ where
\[
\tau_k :=  \inf\{t>\tau_{k-1}: X_t \neq X_{t-} \}, \qquad k \geq 1, \quad\text{with }\tau_0=0,
\]
are the transition epochs, and
\[
R_{k} := X_{\tau_k}-X_{\tau_{k-1}} \in \{\Delta_1,\ldots,\Delta_\bq\}, \qquad k \geq 1,
\]
are the corresponding reactions which may be canonically identified with reaction types $\{1,2,\ldots,\bq\}$. It is immediate that
\begin{align}
X_t = X_0 + \sum_{\tau_k \le t} R_k,
\end{align}
showing the equivalence of the two formulations. We also introduce the counting processes $$N^q_{t} := \sum_{\tau_k \le t} \mathbbm{1}_{\{R_k = \Delta_q\}} \quad\text{and}\quad N^{q,i}_t := \sum_{\tau_k \le t} \mathbbm{1}_{\{R_k = \Delta_q,M_{\tau_k} = i\}}
$$ for the number so far of each reaction by type, so that $$X_t = X_0 + \sum_{q=1}^{\bq} N^q_t \cdot \Delta_q.$$

The probability triple $(\Omega, \F, \PP)$ also supports a finite-state jump process $\{ M_t\}$ that represents the modulating factor. Intuitively, $M_t \in \{1, \ldots, \bam\}$ is a Markov chain with generator
\begin{align}\label{def:Gm}
G_M:= (\mu_{ij}), \qquad i,j \in \{1, \ldots, \bam\}.
\end{align}
Slightly more generally, we take $G_M \equiv G_M(t,X_t)$, allowing the transition rates of $\{M_t\}$ to depend on time and the current state of $\{X_t\}$, so $\{(X_t,M_t)\}$ is jointly Markov.

To complete the description of $\{X_t\}$ and $\{M_t\}$ we finally specify the transition rates $\alpha$ of $\{X_t\}$, or equivalently the arrival rates of the counting process $N^{q,i}$. We assume that the corresponding propensity functions are of the form
\[
\alpha_q(t,X_t,M_t) = \theta_q(M_t) \cdot h_q(t,X_t), \qquad q = 1,\ldots,\bq,
\]
where $\Theta \equiv( \theta_1(1),\theta_1(2), \ldots, \theta_{\bq}(\bam)\}$ are the reaction \emph{rates}, modulated by $\{M_t\}$, and $h$'s are the mass action laws. Let
\begin{align}
\balph(t,X_t,M_t) := \sum_{q=1}^{\bq} \alpha_q(t,X_t,M_t)
\end{align}
be the total current arrival rate, and
\begin{align}
f_q(t,X_t,M_t) := \frac{\alpha_q(t,X_t,M_t)}{\balph(t,X_t,M_t)}
\end{align}
be the conditional  likelihood of the next reaction being of type $q$.

Due to the local Markovian structure, the distribution of the point process $(\tau_k, R_k)$, conditional on a path of $\{M_t\}$ is given explicitly by \citep{AmreinKunsch12}
\begin{align}\label{eq:cond-like-X}
& p \left(\tau_1, R_1, \ldots, \tau_n, R_n | M_s, s\leq t \right) \nonumber
 =  \exp\left(-\int_0^t \balph(s,X_{s},M_s) ds\right)  \\
& \quad \times \prod_{k=1}^n \left( \sum_{q=1}^{\bq} \alpha_q(\tau_k, X_{\tau_k-}, M_{\tau_k}) \cdot \mathbbm{1}_{\{ R_k = \Delta_q \}} \right),
\end{align}
where the first term accounts for the total arrival intensity on $[0,t]$ and the second term for the likelihoods of the observed event types. Recall that if $M_s$ is constant and $\alpha$'s are independent of $t$, then the inter-arrival times are exponentially distributed.

\section{Inference}\label{sec:smc}
Statistical inference in the presented framework consists of estimating the reaction rates $\Theta$ and the seasonal factor $\{M_t\}$. As already mentioned, throughout we assume that $\{X_t\}$ is fully observed, i.e.~by any date $t$, a full record of all reaction times $\tau_k$ before $t$ and corresponding reaction types $R_k$  is available. To explain our method,
 we shall consider the following three cases:
\begin{enumerate}[(a)]
\item $\Theta$ unknown; $\{M_t\}$ observed;
\item $\Theta$ known; $\{M_t\}$ unobserved;
\item Both $\Theta$ unknown and $\{M_t\}$ unobserved.
\end{enumerate}
Case (c) is the main object of interest in our study; however to understand its properties, in the next sections we briefly review Cases (a) and (b).

Probabilistically, the three cases are distinguished by the different filtrations of observed information. Let $(\mathcal{G}_t)$,
\begin{align}
\mathcal{G}_t \equiv \F^{X,M,\Theta}_t :=  \sigma( X_s, M_s : 0 \le s \le t) \vee \sigma(\Theta),
\end{align}
denote the full filtration and using obvious notation let $(\F^{X,M}_t)$, $(\F^{X,\Theta}_t)$, and $(\F^{X}_t)$ denote the sub-filtrations corresponding to cases (a), (b), (c), respectively.
Then our aim is to compute the (joint) posterior distribution $\cX_t := p(M_t, \Theta | \F_t)$
 of the seasonal factor and  the parameters  for the above choices of filtrations.
For later use we also define the posterior probabilities ${\Pi}_t := \left(\Pi_t^1, \ldots, \Pi_t^{\bam} \right)$ where
\begin{align}
\Pi_t^i = & \mathbb{P}^{\mathbf{\pi}}(M_t=i|\mathcal{F}^{X}_t), \qquad i \in \{1, \ldots, \bam \},
\end{align}
where $\mathbb{P}^{\mathbf{\pi}}$ denotes the conditional probability measure given the prior distribution
${\Pi}_0 = \mathbf{\pi}$ of $M_0$.

\subsection{Conjugate Inference of Epidemic Parameters}
If $\{M_t\}$ is observed (i.e.~$(\F^{X,M}_t)$ is available), then $\{X_t\}$ forms a time-inhomogeneous Markov chain. In particular, a full description and analysis of $\{X_t\}$ is possible on the intervals $[\sigma_\ell, \sigma_{\ell+1}]$ where $(\sigma_\ell)$ are the transition times of $\{M_t\}$: $\sigma_{\ell+1} =\inf \{ t \ge \sigma_\ell : M_t \neq M_{t-}\}$. 

It is well known that conjugate Bayesian updating of $\Theta$ is available using Gamma priors \citep{Boys:Wilk:Kirk:baye:2008,AmreinKunsch12}. Specifically, let $$p(\Theta | \F^{X,M}_0) = \prod_{i=1}^{\bam} \prod_{q=1}^{\bq} Ga(\theta_q(i);a^i_{q},b^i_{q}),
$$ where $a_{\cdot}$ and $b_{\cdot}$ denote the shape and rate parameters respectively of the Gamma distribution, be the independent priors of $\theta_q$. Then given $\F^{X,M}_t$ we can express the \emph{full data likelihood} as follows:
\begin{align}\label{eq:conjugate-theta}
p(\Theta|\F^{X,M}_t) = & \prod_{i=1}^{\bam}\prod_{q=1}^{\bq} Ga\left(\theta_q(i); a^i_{q}(t), b^i_{q}(t)\right),\\ \nonumber
 a^i_q(t) := &a^i_{q}+N^{q,i}_{t}, \\ \nonumber
 b^i_q(t) := & b^i_{q}+ \int_0^t  h_q(s, X_s) \mathbbm{1}_{\{ M_s = i\}}\,ds,
	\end{align}
where we recall that $N^{q,i}_t$ is the number of observed reactions of type $q$ during the $i$-regime.

We summarize those sufficient statistics as $\vs_t = (a^i_q(t),b^i_q(t))$, $i=1,\ldots,\bam$, $q=1,\ldots,\bq$ and denote the updating by the function $\mathcal{S}(\cdot)$ from \eqref{eq:conjugate-theta} , i.e.,
\begin{align}\label{def:St}
\vs_T & = \mathcal{S}(\vs_t, t,T, M_{\cdot})  := \Bigl(a^i_q(t)+N^{q,i}_T - N^{q,i}_t, 
b^i_q(t) + \int_t^T h_q(s, X_s, M_s) \mathbbm{1}_{\{ M_s = i\}} \,ds \Bigr)_{q=1}^{\bq}.
\end{align}
Relation \eqref{def:St} provides an explicit sequential way to update the posterior of $\Theta$ along the trajectory of $\{ (X_t,M_t)\}$.

\subsection{Estimation of the Latent Seasonal Factor}
In case (b), we assume that all rates $\Theta$ are known and a full record of $\{X_t\}$ is available, but the path of  $\{M_t\}$ is unobserved. Intuitively, inference of $\{M_t\}$ is based on comparing the likelihoods of observed event epochs/types conditional on the possible values of the seasonal factor, using \eqref{eq:cond-like-X}.
 Using the representation of $\{X_t\}$ as a state-dependent marked doubly-stochastic Poisson process  implies that ${\Pi}_t$ can be described in closed-form and possesses piecewise-deterministic dynamics. Applying \cite[Prop~2.1]{LS07} we obtain the following characterization

\begin{prop} \label{prop:pathPi}
The sample paths of $\{{\Pi}_t\}$ follow
	\begin{equation}\label{eq:pi-flow}
	\left\{
		\begin{array}{ll}\vspace{.2em}
		{\Pi}_t = \vec{x}(t-\tau_k, {\Pi}_{\tau_k}, X_{\tau_k}), \qquad\; \tau_k \leq t < \tau_{k+1}, k\in \mathbb{N}\\
		{\Pi}^i_{\tau_k} =  \dfrac{\alpha_{R_k}(\tau_k, X_{\tau_k-}, i)\Pi_{\tau_k-}^i}{\sum_{j=1}^{\bam} \alpha_{R_k}(\tau_k, X_{\tau_k-}, j)\Pi_{\tau_k-}^j}, \quad i=1,\ldots,\bam, \\
\end{array} \right.
	\end{equation}
where the vector field $\vec{x}(t,\vp,X)$ is defined via \begin{align}
	x^i(t, \vp,X) & = \dfrac{\mathbb{P}(\tau_1>t, M_t=i | M_0 \sim \vp, X_0 = X)}{\mathbb{P}(\tau_1>t | M_0 \sim \vp, X_0 = X)},
	\end{align}
$i \in \{1,\ldots,\bam\}$, and has the explicit solution $\mathbb{P}(\tau_1>t, M_t=i | M_0 \sim \vp, X_0 = X) = \vp \cdot \exp( \int_0^t G_M(s,X)-A(s,X) \,ds)$ with $A(s,X) := \diag( \balph(s,X,\cdot))$ and
\begin{align}\label{eq:Ai}
A(s,X)_{ii'} = \left\{ \begin{aligned}
 \balph(s,X,i) & & \text{if}\; i = i'; \\ 0 & & \text{otherwise}.
\end{aligned} \right.\end{align}
\end{prop}
Proposition \ref{prop:pathPi} completely identifies the distribution of the $\bam$-dimensional posterior $\Pi_t$ of $M_t$ through the recursion \eqref{eq:pi-flow}.

\subsection{Joint Sequential Inference using SMC}

We now turn to our main case (c) where we only have access to $(\F^{X}_t)$, so that neither $\{M_t\}$ nor $\Theta$ is observed. In that case, even if all transitions of $\{X_t\}$ are fully observed, the posterior distribution $\cX_t  = p(M_t,\Theta|\F^X_t)$ no longer admits any sufficient finite-dimensional statistics. Hence, no closed-form analysis is possible and we turn to constructing efficient numerical approximation schemes.

The sequential Monte Carlo approach to recursively update $\cX_t$ consists of constructing a particle approximation
\begin{align*}
\cX_t \simeq \sum_{j=1}^J w_{t}^{(j)}
\delta_{\{m_{t}^{(j)},\theta^{(j)} \}}
\end{align*}
where $\delta$ is the Dirac delta function and each of the $J$ particles is defined by its (normalized) weight $w_t^{(j)} \ge 0$, its $M$-location $m_t^{(j)}$ and its parameter versions $\theta^{(j)}$. In other words,
$$
\PP( (M_t, \Theta) \in A | \F^X_t) \simeq \sum_{(m_t^{(j)},\theta^{(j)}) \in A} w_t^{(j)}.
$$
The particles are updated using a propagation-selection scheme. However, since the parameters $\Theta$ are fixed throughout,
it is well known that naive SMC implementation typically leads to \emph{particle degeneracy}, namely the diversity of $\theta^{(j)}$'s across the particles increasingly diminishes due to resampling. As we shall see in Section \ref{sec:compare}, this issue is acute for HMSKMs.

The availability of the sufficient statistics $\vs_t$ from \eqref{def:St} for $\Theta$ conditional on $\{M_t\}$ allows to dramatically reduce degeneracy. We recall the early approach of
\cite{Stor:part:2002} who applied a basic bootstrap particle filter \citep{Gord:Salm:Smit:nove:1993} on the pair $(X_t,\vs_t)$, where the likelihood of observations is evaluated by sampling from the posterior of $\Theta$, i.e.
\begin{align}\label{eq:storvik}
p(X_t | M_{t}, \vs_t) \simeq p(X_t|M_t, \theta^{(j)}) \qquad \theta^{(j)} \sim \vs_t \quad\text{i.i.d.}
\end{align}

\subsection{Particle Learning in HMSKM}

In our case, even more efficiency can be achieved through a resample-move SMC \citep{Gilk:Berz:foll:2001} which takes advantage of the explicit predictive likelihood of $\cR$ given $\Theta$ (see \eqref{eq:pred-like} below) and leads to a version of the {particle learning} (PL) framework originally proposed within a discrete-time setting in \cite{CarvalhoPL10,CarvalhoPL11}. The use of a \emph{resample-move} algorithm allows direct sequential  filtering of $(M_t, \vs_t)$ rather than the static $\Theta$. Thus, the  filtered distribution at time $t$ is approximated by the particle cloud
\begin{align}
\cX_t \simeq \cX^{(J)}_t := \sum_{j=1}^J
w^{(j)}_{t}
f_\Theta(\cdot |s_{t}^{(j)}) \delta_{\{m_{t}^{(j)},s_{t}^{(j)}\}},
\end{align}
where $s_{t}^{(j)}$ are again the parameter sufficient statistics in \eqref{def:St}. Thus, the conditional distribution of $\Theta$, $f_\Theta(\cdot | \vs_t)$, is a product of independent Gamma's and
the overall ${\cX}^{(J)}_t$ represents the posterior of each $\theta_q$ as a mixture of Gamma distributions.

\begin{algorithm}[H]
\caption{Particle Learning for Hidden Markov Stochastic Kinetic Model\label{algo:PL}}
{\fontsize{10.5}{13}
\begin{algorithmic}[1]
\REQUIRE{ Priors $\vs_0$, ${\Pi}_0$, initial state $X_0$, number of particles $J$}
\STATE Sample $m_0^{(j)} \sim {\Pi}_0$ i.i.d., set $s_0^{(j)} \leftarrow \vs_0$, $j=1,\ldots,J$
\LOOP[ for $k=0,1,\ldots,$] 
\STATE Sample $\theta^{(j)} \sim p(\theta|s_{\tau_k}^{(j)})$, $j=1,\ldots,J$
\STATE  Calculate weights $w_{k+1}^{(j)} \propto p(\tau_{k+1}-\tau_k,R_{k+1} |m_{\tau_k}^{(j)},\theta^{(j)})$ 
\FOR{$j=1,\ldots,J$}
\STATE Re-sample $j' \propto w_{k+1}^{(\cdot)}$ where $j' \in \{1,\ldots,J\}$
\STATE Sample a trajectory $m_{(\tau_k,\tau_{k+1}]}^{(j)}$ using the conditional law $p( M_{\cdot} |m_{\tau_k}^{(j')},\theta^{(j')}, \tau_{k+1}, R_{k+1})$
\STATE Update $s_{k+1}^{(j)} \leftarrow  \mathcal{S}\bigl(s_{k}^{(j')}, \tau_k, \tau_{k+1}, m_{(\tau_k,\tau_{k+1}]}^{(j)}\bigr)$
\ENDFOR
\ENDLOOP
\end{algorithmic}}
\end{algorithm}

Algorithm \ref{algo:PL} summarizes in pseudo-code the steps of the proposed particle learning algorithm. Its main steps are computing particle weights in (4) for resampling using the predictive likelihood of the next event, forward propagation step (6) using the conditional law of the environment factor, and updating of the sufficient statistics step (7) for the parameters. Overall, besides the analytical results detailed below, only the ability to simulate $\{M_t\}$ and $\{X_t\}$ is needed to implement the above Monte Carlo scheme, highlighting the flexibility of PL. A basic simulation method for SKM that is exact and can always be used is the Gillespie algorithm (see e.g.~ \cite{Wilk:stoc:2006}), here slightly extended to take into account additional transition times of $\{M_t\}$.

To calculate the \emph{predictive likelihood} of the next inter-arrival interval $\tau_{k+1}-\tau_k$ and reaction type $R_{k+1}$ conditional on $M_{\tau_k}$ and parameters $\Theta$ we rely on the analytic expression in \eqref{eq:cond-like-X},
	\begin{align}
	 p(\tau_{k+1}-\tau_k,R_{k+1} |M_{\tau_k},\Theta)  =   \sum_{i=1}^{\bam} p( R_{k+1} |M_{\tau_{k+1}}=i, \Theta)  
 \label{eq:pred-like}  & \times p(\tau_{k+1}-\tau_k,M_{\tau_{k+1}}=i|M_{\tau_k}, \Theta).
	\end{align}
	The first term on the right-hand-side is  $f_r(\tau_{k+1}, X_{\tau_{k+1}-},M_{\tau_{k+1}})$ using the parameters $\Theta$,
	and the second term is
	\begin{align}
	\PP(\tau_{k+1}-\tau_k=t, M_{\tau_{k+1}}=i|M_{\tau_k}=i^{\prime}, \Theta) 
:= P_{i^{\prime}i}(t) \balph(\tau_{k+1}, X_{\tau_{k+1}-},i)
	\end{align}	
	with $P_{i^{\prime}i}(t)$ being an element of the matrix exponential $P(t) = e^{\int_0^t G_M(s,X) - A(s,X) ds}$, see \eqref{eq:Ai}. When there are just two latent states, $|\bam|=2$, $P(t)$ can be computed explicitly using eigenvector decomposition.

The \emph{conditional law} of $\{M_t\}$ given $\tau_{k+1}-\tau_k,R_{k+1}$ is not available in closed form. However, using Bayes rule
\begin{align}\nonumber
p & (M_{(\tau_k,\tau_{k+1}]}|M_{\tau_k},s_{\tau_k},\Theta, \tau_{k+1},R_{{k+1}})
	\\ & \propto 
     p(\tau_{k+1}-\tau_k,R_{k+1}|M_{(\tau_k,\tau_{k+1}]},\Theta)p(M_{(\tau_k,\tau_{k+1}]}|M_{\tau_k}) \label{eq:cond-like}
	\end{align}
	where
	\begin{align}\label{eq:cond-law-M}
	p&(\tau_{k+1}-\tau_k,R_{k+1}|M_{(\tau_k,\tau_{k+1}]},\Theta) 
	\\  \nonumber &= \exp\left\{-\int_{\tau_k}^{\tau_{k+1}} \!\balph(s,X_{\tau_k},M_s)\, ds\right\} \cdot \alpha_{R_{k+1}}(\tau_{k+1},X_{\tau_k},M_{\tau_{k+1}})
	\end{align}
    and $p(M_{(\tau_k,\tau_{k+1}]}|M_{\tau_k})$ is determined from the transition matrix of $\{M_t\}$. To implement \eqref{eq:cond-like} we use a rejection sampling step relying on the fact that there is an easy upper bound of \eqref{eq:cond-law-M} :
      \begin{multline}\label{eq:max-like-M}
        p(\tau_{k+1}-\tau_k,R_{k+1}|M_{(\tau_k,\tau_{k+1}]},\Theta)  \\  \le \exp\!\bigl\{(\tau_{k}-\tau_{k+1}) \min_{i,s} \balph(s,X_{\tau_k},i)\, \bigr\} \! \cdot \max_i \alpha_{R_{k+1}}(\tau_{k+1},X_{\tau_k},i).
            \end{multline}
      Thus, we simulate using the unconditional law $p(M_{(\tau_k,\tau_{k+1}]}|M_{\tau_k})$ and then accept the simulation with probability given by the ratio between \eqref{eq:cond-law-M} and \eqref{eq:max-like-M}.
    Since $\tau_k$ are typically tightly spaced, the above conditional likelihoods are all close to each other, requiring only a few additional simulations (acceptance probability is usually $>95\%$).

Finally, as already mentioned, the sufficient statistics for $\Theta$ are always conjugate-Gamma with the updating given explicitly in \eqref{def:St}. We note that for typographical convenience in Algorithm \ref{algo:PL} resampling takes place at each reaction time $\tau_k$. In practice, any other resampling frequency can be chosen; in that case the weights are updated accordingly until resampling takes place. Also, a variety of resampling schemes (multinomial, residual, stratified, etc.) are available \citep{RydenBook} and can be used to lower Monte Carlo error.

Like all SMC methods, PL still exhibits sample impoverishment which implies that the filtering error $\| \cX_T^{(J)} - \cX_T \|$ (in an appropriate metric) grows exponentially in $T$. Thus, exponentially more particles are needed to control the Monte Carlo error in terms of the number of observations. We therefore recommend utilizing PL on a fixed horizon $T$ as in our application below.

\section{Model of Seasonal Epidemics}\label{sec:sis}
We now return to our main example of a Markov-modulated chemical reaction system --- a compartmental model of seasonally-forced endemic diseases. As a simple example, we  shall analyze a classical stochastic SIR-type model \eqref{eq:reactions} of epidemics that incorporates a latent seasonal factor $\{M_t\}$. For concreteness, we phrase our discussion in terms of the human influenza virus.

A basic description of an endemic disease such as flu can be provided using an SIS compartmental model \citep{AnderssonBrittonBook}, which features just two population compartments of Susceptibles $\{ S_t\}$ and Infecteds $\{ I_t\}$. Since only partial immunity is available against influenza (it is common for an individual, especially children, to have several flu episodes in one season), the Recovered compartment is omitted and we assume that upon recovery individuals immediately pass back into the susceptible pool. Other models of endemic diseases are reviewed in \cite{Nasell02}. 

Let $M_t \in \{1,2\}$ denote the seasonal factor at date $t$, with $M_t =1$ representing low flu  season and $M_t = 2$ high flu season. We assume a closed population of constant size $N := S_t + I_t$ that represents a fixed geographic area (such as a college campus, a town, or a county). We assume that $\{M_t\}$ forms a time-stationary Markov chain with infinitesimal generator
$$
G_M := \begin{pmatrix}
-\mu_{12} & \mu_{12} \\
\mu_{21} & -\mu_{21}
\end{pmatrix}.
$$
In other words, holding times for $\{M_t\}$ in regime $i$ are exponentially distributed with mean $\mu_{ij}^{-1}$.
Conditional on $M_t = i$, $\{X_t\} \equiv \{S_t,I_t\}$ is a jump-Markov process involving two reaction types
\begin{align}\label{eq:reactions}
\left\{ \begin{aligned}
\text{Infection:} & & S + I & \xrightarrow{\theta_1(i) h_1} 2I &   \; h_1 := ( I_t + \imm)\frac{S_t}{N}; \\
\text{Recovery:} & & I & \xrightarrow{\theta_2(i) h_2} S &  \quad h_2 := I_t. \\
\end{aligned} \right.
\end{align}
Thus, new infections take place according to the law of mass action \citep{AnderssonBrittonBook}, where the infection rate is driven by the possible pairings between infected and susceptible individuals (assuming homogenous mixing of the full population). Additionally, we add an ``immigration of infecteds'' rate $\imm$ which can be viewed as an external reservoir of the flu (e.g.~from travellers) that provides a constant source of additional infections. We use this term to prevent a stochastic fade-out of the epidemic and guarantee endemicity. There are two epidemic parameters $\Theta \equiv (\theta_1,\theta_2)$, interpreted as infectiousness ($\theta_1$) and mean recovery time ($1/\theta_2$). Since $S_t = N-I_t$, $\{I_t\}$ summarizes the epidemic state and we omit $S_t$ from further discussion.

For simplicity, we assume that seasonal variations affect only the contact rate $\theta_1(M_t)$, so that $\theta_2(M_t)=\theta_2$ is constant. We moreover assume that the effect on the contact rate is multiplicative, $$\theta_1(2) = (1+SF)\theta_1(1),
 $$
for a known seasonality impact ratio $SF > 0$.
 This is meant to model the case where the seasonality increases probability of susceptibles becoming infected (due e.g.~to cold weather) but has no impact on the severity of the flu once infected.

As described above, given $M$, the key state $\{I_t\}$ forms a recurrent
 Markov chain on $\{0,1,\ldots, N\}$.
Using the notation in Section \ref{sec:general}, we have $R_k \in \{-1,1\}$ with
\begin{align}
\alpha_1(I,i) := \theta_1(i)\frac{(I +\imm) (N-I)}{N}, \quad \alpha_{-1}(I,i) := \theta_2 I,
\end{align}
for $i=1,2$, and \begin{align}
f_r(I, i) = \left\{
	\begin{array}{rl}
	\frac{ \theta_1(i) (I + \imm) (N-I)/N }{ \theta_1(i) (I+\imm)(N-I)/N +\theta_2 I } & \quad\text{if}\quad r=1; \\
	\frac{\theta_2 I}{\theta_1(i) (I+\imm)(N-I)/N +\theta_2 I} & \quad\text{if}\quad r=-1.
	\end{array} \right.
\end{align}

Since $\{M_t\}$ only takes on two values, its posterior is described by the one-dimensional probability process
$$
\Pi^2_t := \PP( M_t = 2 | \F^{I}_t).
$$
Applying Proposition \ref{prop:pathPi} and using the fact that $\theta_2$ is independent of $\{M_t\}$ leads to the following simple dynamics of $\Pi^2_t$:
\begin{cor} \label{prop:pathPi-SIS}
The evolution of $\{\Pi^2_t\}$ is characterized on each $t \in [\tau_k,\tau_{k+1})$ by
	\begin{align}\label{eq:riccati-sis}
		 \frac{d\Pi^2_t}{dt}  = & (1- 2 \Pi^2_t)\mu_{21}  
 - SF \cdot \theta_1 \frac{(N-I_{\tau_k})(I_{\tau_k}+\imm)}{N}\Pi^2_t(1-\Pi^2_t),\\
		\text{and} \; \Pi^2_{\tau_k}
&=\left\{
\begin{array}{rl}
\frac{(1+SF)\Pi^2_{\tau_k-}}{1+SF \Pi^2_{\tau_k-}} > \Pi^2_{\tau_k-}, & \text{if }\quad R_{k}=1, \\
\Pi^2_{\tau_k-},  &\text{if } \quad R_{k}=-1. \end{array}\right\}.
	\end{align}
An explicit solution to \eqref{eq:riccati-sis} can be obtained by computing the eigen-pairs of the corresponding matrix $A(I)$.
\end{cor}

The particle learning Algorithm  \ref{algo:PL} can be straightforwardly applied in this model using particles of the form
$(m_t^{(j)},s^{(j)}_t) \in \{1,2\} \times \R_+^4$. Since $\theta_2$ is not modulated, its posterior is in fact \emph{deterministic} given $\F^I_t$, i.e.~the same for all particles and only the two sufficient statistics of $\theta_1$ need to be recorded:
\begin{align*}
p \left(\theta_1 | \F^I_t, \{m_s^{(j)}, 0 \le s \le t\} \right) & = Ga \left( a_1 + N^1_t, b_1(t) \right); \\
p\left(\theta_2 | \F^I_t \right) & = Ga\bigl( a_2 + N^2_t, b_2 + \int_0^t I_s \,ds \bigr) \!;\\
 b_1(t) := b_1 + \int_0^t  \Bigl(1+ SF & \mathbbm{1}_{\{ m_s^{(j)} = 2\}} \Bigr)\frac{(I_s + \imm)(N-I_s)}{N} \,ds,
\end{align*}
where $\vs_0 = (a_1,a_2,b_1,b_2)$ is the original Gamma prior of $\Theta$.

\subsection{Illustration}\label{sec:illustrate}
To illustrate the above algorithms, this section presents several numerical experiments. Table \ref{tbl:params} summarizes the parameters used. We consider a single flu season  lasting 9 months (approximately September through May), which includes on average a 6-month long high-season (so that $\mu_{21}^{-1} = 365/2$).  Such a model is meant to capture a single seasonal cycle as is commonly done by US epidemiological agencies on a Fall-Spring basis, see Figure \ref{fig:seasonal-flu}.

We consider a fixed population of $N=10,000$ individuals with an initial $I_0 = 50$ infecteds. During the low season, $\theta_1(1) < \theta_2$ meaning that the epidemic would on its own fade out; the disease remains endemic through contact with outside infecteds, with a carrying capacity (i.e.~the long-run expected level of Infecteds freezing environmental fluctuations) of about $\lim_{t\to\infty} \E[ I_t | M_s = 1 \forall s] \simeq 50$. In the high season, $\theta_1(2) > \theta_2$ so that an outbreak begins. Even though the infectiousness rate increases by just $SF = 15\%$, the resulting carrying capacity  jumps to over $500$, i.e.~more than 5\% of total. This illustrates that even small changes in the contact rate can have a dramatic effect on the equilibrium disease incidence. In line with common estimates, we use average infectiousness period of about 4 days, $\theta_2 \simeq 0.25$.

\begin{table} [h]
\caption{\label{tbl:params} Parameter values used. All rates are daily.}
\centering{\begin{tabular}{@{}llr@{}} \hline
Parameter & Meaning & Value \\ \hline \hline
$\mu_{12}$ & \text{Transition rate to high season} & 6/365 \\
$\mu_{21}$ & \text{Transition rate to low season} & 2/365 \\
$\theta_2$ & \text{Recovery rate} & 0.25 \\
$\theta_{1}(1)$ & \text{Low-season infectiousness} & 0.235 \\
$\theta_{1}(2)$ & \text{High-season infectiousness}  & 0.27025 \\
$SF$ & \text{Seasonality effect} & 0.15 \\
$\imm$ & \text{Immigration of outside infecteds} & 2 \\
$N$ & \text{Population size} & 10,000 \\
$T$ & \text{Time horizon (days)} & 273 \\
$(M_0,I_0)$ & \text{Initial Condition} & (0,50) \\
$(a_1(0),b_1(0))$ & \text{Initial Priors for $\theta_1$} & (25,100) \\
$(a_2(0),b_2(0))$ & \text{Initial Priors for $\theta_2$} & (25,100) \\
 \hline\hline
\end{tabular}}
\end{table}

\begin{figure}
\begin{center}
\includegraphics[width=0.59\textwidth,trim=0.35in 2.5in 0.3in 2.5in]{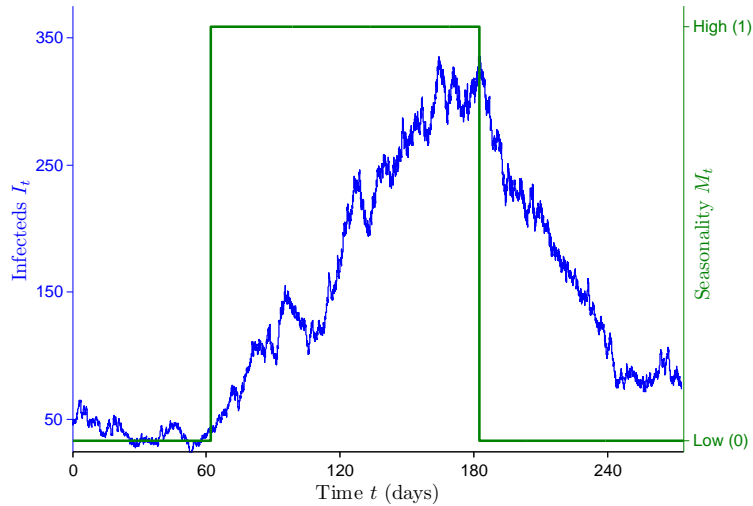}
\caption{Sample path of $\{M_t, I_t\}$ over a course of 9 months. There are a total of $20079$ transitions.}\label{fig:IvsX}

\end{center}
\end{figure}

Figure \ref{fig:IvsX} shows a sample trajectory of the infected population count over the nine months in conjunction with the underlying seasonal factor $\{ M_t\}$. In this scenario, starting with low-season, high season $\{M_t = 2\}$ begins on day 61 and ends on day 182, lasting just over four months (compared to average high season length of 1/2-year). The seasonality effect is clear, as soon after the beginning of the high season, $\{I_t\}$ begins an upward trend which is reversed once $M_t = 1$ again. Nevertheless, we observe a lot of stochastic fluctuations against these main trends (e.g.~a significant decrease in $I_t$ around day 100).  Overall, there were a total of $10051$ infections recorded over the period, corresponding to roughly each member of the population becoming infected once, in line with observed statistics on influenza.  Peak number of infecteds was 335 on day 164.

\begin{figure}
\begin{center}
\hspace*{-0.15in}\includegraphics[height=2.2in,trim=1.7in 3.6in 1.7in 3.5in]{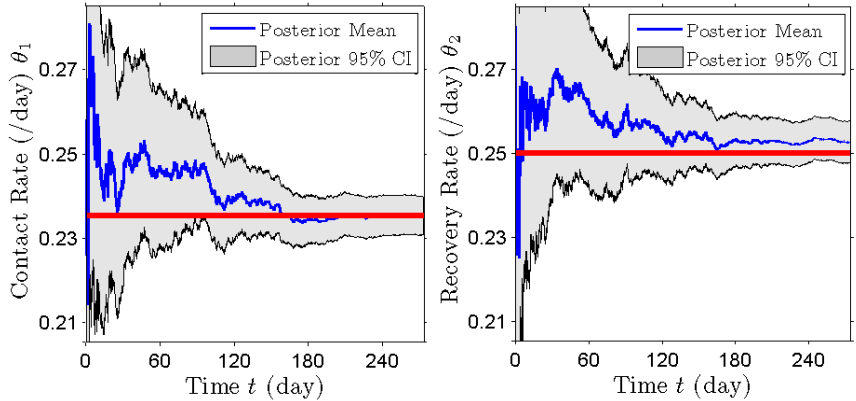} 
\caption{Posterior mean and 95\% credibility interval of $\Theta \equiv (\theta_1, \theta_2)$ over the sample path of $\{M_t, I_t\}$ in Figure \ref{fig:IvsX}. Solid horizontal lines indicate the true values used. }\label{fig:paramPost}

\end{center}

\end{figure}

Starting with a rather vague prior for $\Theta$, Figure \ref{fig:paramPost} shows the sequential Bayesian inference of $\Theta$ assuming that the trajectory of $\{M_t, I_t\}$ shown in Figure \ref{fig:IvsX} is fully observed. We note that the posterior means apparently converge to the true values and the posterior credibility interval (CI) narrows at roughly a hyperbolic rate over time. As data is accumulated, the oscillations in the posterior distributions decrease quickly.

\begin{figure}[ht]
\begin{center}
\includegraphics[width=0.59\textwidth,trim=0.3in 2.2in 0.25in 2.9in]{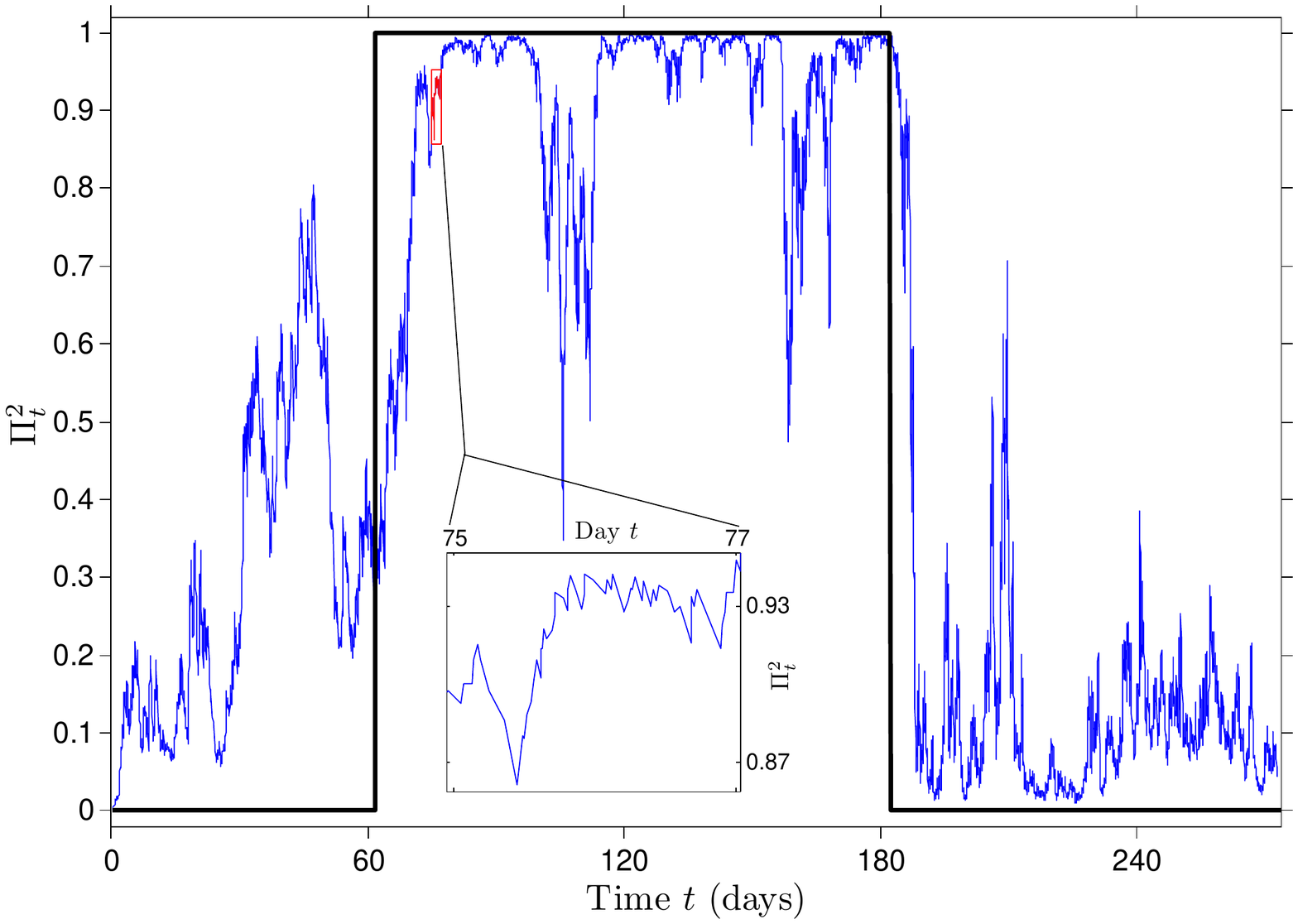}
\caption{Posterior probability $\Pi^2_t = \PP ( M_t = 2 | \F^I_t)$ over the sample path of $\{I_t\}$ in Figure \ref{fig:IvsX}, assuming known $(\theta_1,\theta_2)$. For comparison we also show the respective true trajectory of $\{M_t\}$.
}\label{fig:PitPath}

\end{center}

\end{figure}

Figure \ref{fig:PitPath} presents the results from the other nested model (b), namely assuming known parameters $\Theta$ but unobserved seasonal factor $\{M_t\}$. In Figure \ref{fig:PitPath} we show the posterior probability $\Pi^2_t$ of the high season over the same trajectory of $\{I_t\}$ shown in Figure \ref{fig:IvsX}. We note that generally the filter is able to well-identify the present seasonality effect and responds very quickly when $\{M_t\}$ changes (see the very sharp drop around $t=185$). The dynamic lag between change in the true $\{M_t\}$ and the response by the filter is on the order of 10-20 days which is quite fast and would be difficult to identify with a ``naked eye'' looking at the trajectory of $\{I_t\}$. At the same time, the filter $\Pi^2_t$ is highly sensitive to the local behavior of $\{I_t\}$ making it very noisy. For example the mentioned drop in infecteds around $t=100$ causes posterior likelihood of high season to decrease from over 95\% to as low as 35\%, albeit with a sharp reversal once the upward trend is re-established. The underlying piecewise-deterministic behavior of $\{\Pi^2_t\}$ (see Proposition \ref{prop:pathPi}) is highlighted in the inset figure which clearly shows the discrete upward jumps of $\{\Pi^2_t\}$ when new infections are recorded.

\begin{figure*}
\begin{center}
\hspace*{-0.2in}\includegraphics[height=2.5in,trim=0.35in 3.9in 0.35in 3.4in]{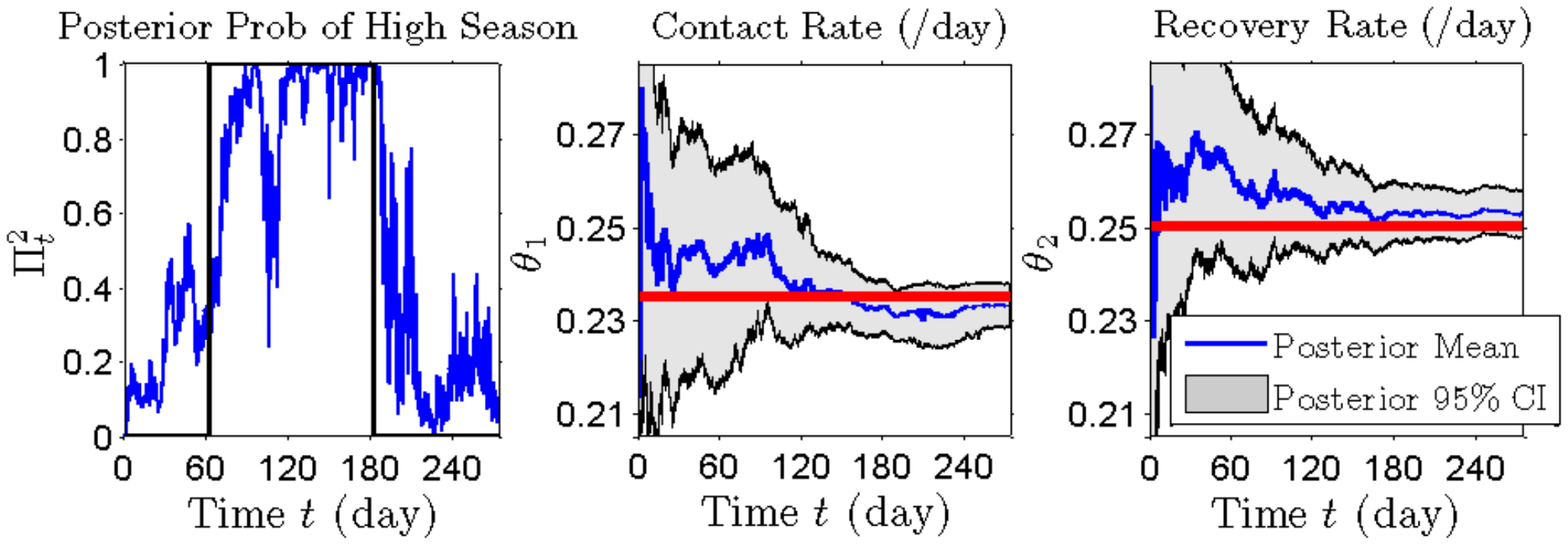}
\caption{Joint inference of $(M_t,\Theta)$ over the trajectory of $\{I_t\}$ in Figure \ref{fig:IvsX}. Left panel shows the posterior probability of the high season $\Pi^2_t = \PP( M_t = 2 | \F^I_t)$. The other two panels show the posterior median and 95\% credibility interval of the two parameters $\theta_1, \theta_2$. The SMC algorithm used $J=5000$ particles. All other parameters are from Table \ref{tbl:params}.
}\label{fig:jointPost}
\end{center}

\end{figure*}

Finally, Figure \ref{fig:jointPost} shows the output from running the SMC algorithm for joint inference of $(M_t, \Theta)$ using $J=5000$ particles. Uncertainty about both parameters and seasonal factor makes the posterior credible intervals wider compared to Figure \ref{fig:paramPost}. In terms of the posterior probability of the high season, the resulting $\Pi^2_t$ is less volatile compared to Figure \ref{fig:PitPath} and responds slightly slower to underlying regime shifts.

The implemented instance of Algorithm \ref{algo:PL} is somewhat computationally intensive since it requires repeated simulation of paths of $\{M_t\}$ and evaluation of the predictive and conditional likelihoods over each interval $[\tau_k,\tau_{k+1})$ (so over $20,000$ times in the Figures shown) and for each particle $(J=5000)$. However, we note that all these computations are \emph{exact}, so the only noise present is from Monte Carlo re-sampling. Running time to generate Figure \ref{fig:jointPost} is about two minutes on a typical 2011 laptop.

\subsection{Comparison to Other Inference Methods}\label{sec:compare}
From the time-series estimation point of view, the stochastic systems we consider are characterized by long time-series and short inter-event periods. It is well known that the basic challenge of Sequential Monte Carlo over long horizons is particle degeneracy since repeated re-sampling necessarily throws some information away and cuts down on diversity of the empirical particle cloud. Degeneracy is exacerbated when one is required to estimate constant parameters. Without further steps, the basic bootstrap filter of \cite{{Gord:Salm:Smit:nove:1993}} will  degenerate, almost surely as $t \to \infty$, to a point mass estimate of the posterior  density. A popular simple solution is the \cite{LiuWest} algorithm (henceforth LW) that introduces adjustment moves to particle versions $\theta^{(j)}$ of $\Theta$. We implemented LW as a comparison to the presented PL algorithm and found that degeneracy is still prevalent over the second half of the season even with as many as $J=5000$ particles. As such, use of the sufficient statistics $\mathbf{s}_t$ for $\Theta$-posteriors is \emph{crucial} for HMSKMs.

We also implemented the \cite{Stor:part:2002} filter. Compared to the PL filter, its main difference is that  \cite{Stor:part:2002} applies propagate-resample steps, and as such does not require analytic form of the predictive likelihood \eqref{eq:cond-like} in Step 4 of Algorithm \ref{algo:PL}. Instead, one simply uses
$$
w^{(j)}_{\tau_k} \propto  \exp \left( - \int_{\tau_{k-1}}^{\tau_k} \balph(I_s,m^{(j)}_s) \,ds \right) \cdot \alpha^{(j)}_{R_k}(I_{\tau_k-},m^{(j)}_{\tau_k}),
$$
where the propensity rates $\alpha$'s use the sampled particle versions $\theta^{(j)} \sim \vs_{\tau_{k-1}}$. The Storvik algorithm runs slightly faster since it does not need to evaluate the matrix exponential in \eqref{eq:riccati-sis}.

\begin{figure}
\begin{center}
\hspace*{-0.1in}\includegraphics[width=0.45\textwidth,trim= 0in 0.35in 0in 0.35in]{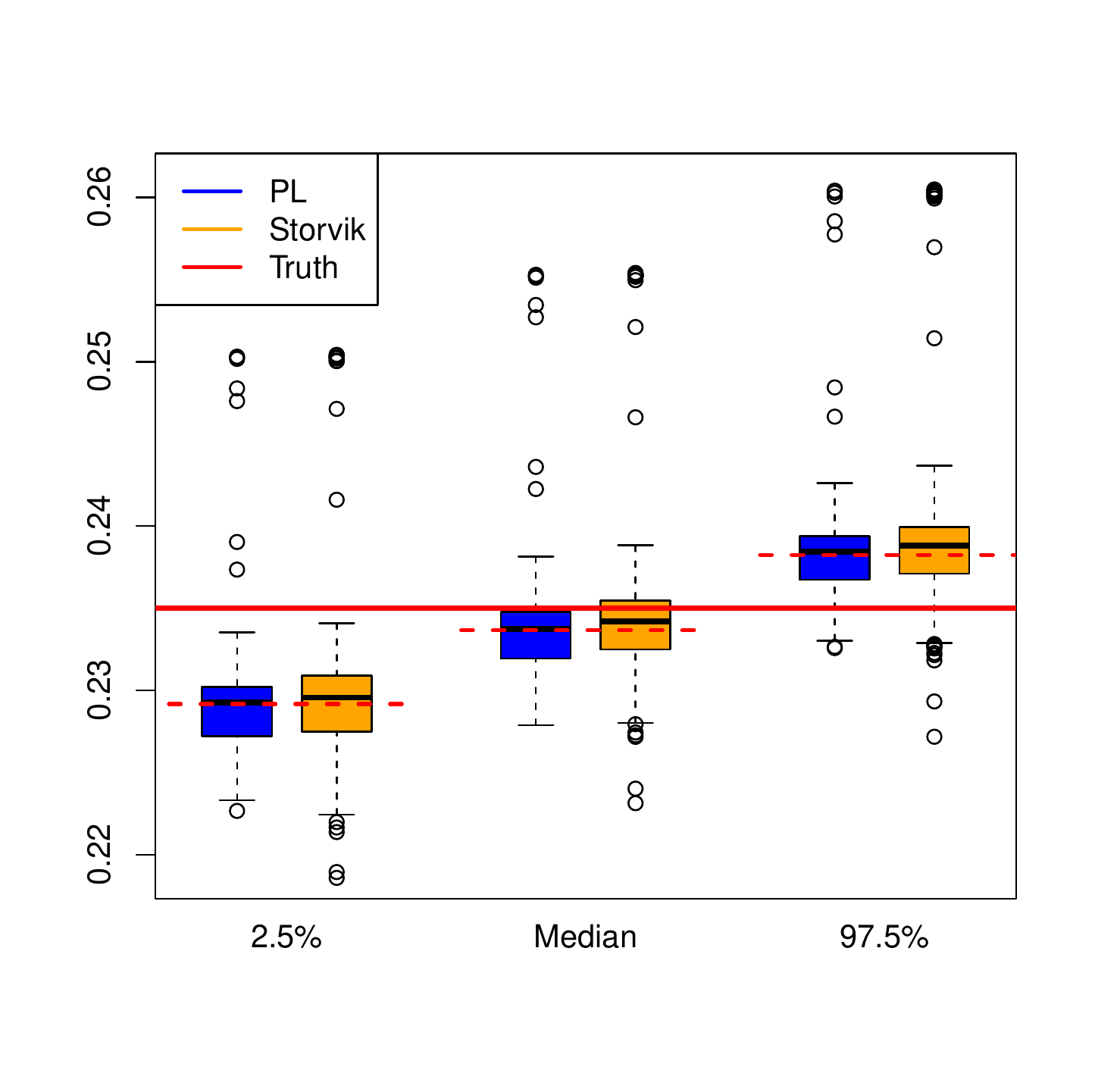}
\caption{Posterior quantiles of $\theta_1$ at $T_2 = 270$ days using the PL and Storvik algorithms with $J=2000$ particles. The histograms show the median and 95\% CI quantiles across 100 runs of each algorithm. The true value of $\theta_1$ is indicated by the solid horizontal line, the dashed lines indicate the corresponding quantiles from a benchmark run of PL with $20000$ particles.
}\label{fig:compare}
\end{center}

\end{figure}

To quantitatively compare the performance of the described SMC algorithms, we started by creating a benchmark using a PL filter with $2 \cdot 10^4$ particles. We then generated 100 runs of the PL, Storvik and LW schemes each using $J=2000$ particles and residual resampling for the realization of $\{M_t,I_t\}$ in Figure \ref{fig:IvsX}. Because in the case study the second reaction rate $\theta_2$ is independent of $\{M_t\}$, its sufficient statistic $\vs^2_t$ is the same for all particles and is computed exactly by both Storvik and PL algorithms for any filter size $J$. On the other hand, the estimate of $\theta_1$ is highly sensitive to correct tracking of $\{M_t\}$ over time. We first compare the $95\%$ coverage probabilities with respect to the true  $\theta_1$, i.e.~how frequently does the true value $\theta_1=0.235$ belong to the corresponding 95\% posterior CI obtained from SMC. The results for the two time-points of $T_1=120$ and $T_2 = 270$ days are summarized in Table \ref{tbl:coverage}. We find that the PL algorithm performs best; the Storvik algorithm has somewhat higher MC errors but still maintains particle diversity. The LW algorithm posteriors start to collapse mid-way through the data and completely degenerate by the end, so that their coverage probabilities are nil.
Figure \ref{fig:compare} further shows that the posterior 95\% CI of the PL algorithm includes fewer outliers, in other words nearly all runs of the algorithm recover the correct posterior.

In Table \ref{tbl:coverage} we also
compare the standard error $\E \left[\| \Pi^2_t - \hat{\Pi}^2_t \|_2 \right]$ of the posterior probability of the high season $\Pi^2_t$ with respect to the benchmark filter $\hat{\Pi}^2_t$. We observe that compared to PL, the Storvik scheme is more prone to ``losing track'' of $M_t$, in other words the Monte Carlo runs appear to be more leptocurtic. The LW algorithm in fact usually tracks $\hat{\Pi}^2_t$ well but also had several completely failed runs and clearly suffers from particle degeneracy.  Overall, this analysis confirms our preference for PL; its advantages can be compared to the improvement from the bootstrap to the auxiliary particle filter in the classical SMC setup.

\begin{table} [h]
\caption{\label{tbl:coverage} Comparison of SMC algorithm performance. All algorithms used $J=2000$ particles and we took $T_1=120$, $T_2=270$ days for the path of $\{M_t, I_t\}$ in Figure \ref{fig:IvsX}. The LW algorithm had a tuning parameter of $h=0.97$.}
\centering{\begin{tabular}{@{}lllr@{}} \hline
 & PL & Storvik & LW \\ \hline \hline
95\% Cov. Prob of $\theta_1$ at $T_1$ & 0.96 & 0.87 & 0.32 \\
95\% Cov. Prob of $\theta_1$ at $T_2$ & 0.74 & 0.65 & 0.01 \\
Std Error of $\Pi^2_t$ at $T_1$ & 0.200 & 0.272 & 0.455\\
Std Error of $\Pi^2_t$ at $T_2$ & 0.056 & 0.105 & 0.037\\
 \hline\hline
\end{tabular}}
\end{table}

\subsection{Discussion}\label{sec:discuss}
The presented model is clearly stylized and would not be able to capture all the features of real epidemics. Since both $\Theta$ and $\{M_t\}$ are assumed unobserved, it may still be possible to fit real data even under such model mis-specification. Nevertheless, in this section we discuss some of adjustments that may be made for achieving further realism.

While year-to-year epidemics arise in different times, there are clear patterns which imply that assuming time-stationary transition rates of $\{M_t\}$ is not reasonable. Our general setup allows for $G_M(t)$ in \eqref{def:Gm} to be time-dependent and could be used to capture these patterns. Other calendar-year effects, such as the impact of the school-year start \citep{IonidesKingMeasles10}, can be deterministically added. A more serious concern is the assumption of known seasonal transition rates $\mu_{ij}$; however in our 2-state example those may be reasonably well-calibrated using historical data such as that used in Figure \ref{fig:seasonal-flu}. Our numerical experiments also show that mis-specification of $\mu_{ij}$ is not a serious problem for sequential inference of ${\Pi}_t$.

The model \eqref{eq:reactions} assumes that there is a single infectiousness parameter $\theta_1$ which remains constant throughout, such that the actual contact rates are of the form $\alpha_1(I_t, M_t) = \theta_1 (1+ SF \mathbbm{1}_{\{M_t = 2\}}) h_1(I_t)$. This representation is for convenience only; it is straightforward to consider the case where we separately carry along priors for each $\theta_1(i)$. We could also consider the case where each new season leads to a \emph{``fresh''} $\theta_1$ (e.g.~from a new strain of the pathogen), i.e.~the contact rate on each interval $[\sigma_\ell, \sigma_{\ell+1})$, where $(\sigma_\ell)$ are the transition epochs of $\{M_t\}$ is $\theta_{1,(\ell)}(M_{\sigma_\ell}) \sim p(\theta)$. In that case, the  sufficient statistics $\vs_t^{(j)}$ are simply reset to the original prior when the corresponding particle copy of $m_{t}^{(j)}$ changes states. However, the difficulty with this method is that $\theta_1(i)$'s are not independent. For instance, to have the interpretation of $M_t=2$ being the high season, we require $\theta_1(2) > \theta_1(1)$. This precludes the simple specification of the respective marginals through a Gamma prior. The assumption of a constant seasonal factor $SF$ is a convenient work-around which enforces the epidemiological meaning of seasonality.

Our choice of a two-compartment SIS model was due to its simplicity. Since our methods can handle any HMSKM, it is immediate to extend to further compartments (e.g.~an SEIRS model with $X_t = (S_t, E_t, I_t, R_t)$ that would further include the Exposed compartment of individuals who are infected but not infectious, and the Recovered compartment for individuals who have temporary immunity) or multiple seasonal regimes (e.g. adding a third ``pandemic'' regime to capture outbreaks like 2009 H1N1 influenza). By modifying the propensity functions $h_q(t,X_t)$ one can also refine the modeling of population mixing \citep{IonidesKingMeasles10}, incorporate external immigration of individuals, or include age-structured populations.

Finally, an extra possibility is to allow two-way feedback between the epidemic state $\{X_t\}$ and the seasonal factor $\{M_t\}$. Thus, beyond the modulation of the transition times of $X$ by $M$, we can also introduce effect of $X$ on $M$ through $G_M = G_M(t,X_t)$. In other words, the seasonality dynamics are themselves affected by the epidemic. For instance, a large outbreak could be due to a long-surviving pathogen which in turn prolongs the expected length of a high-season. Alternatively, one could consider the case where each infection increases the chances of a genetic mutation of the pathogen, thereby decreasing $\mu_{21}$. We refer to \cite{LudkovskiCDC12} for SMC algorithms to address this possibility.

The most severe constraint of our model is the assumption of fully observing $\{I_t\}$. In principle this could be achieved by exhaustive monitoring of all individuals' status. More realistically, we view this model as an idealized case of biosurveillance
 which contends with extrinsic model uncertainty regarding $\Theta$ and $\{M_t\}$ while eschewing missing data. This then provides a useful benchmark to analyze data collection quality. We refer to \cite{LN11} for a related setup with discrete-time observations based on binomial sub-sampling which also admits a PL algorithm.

\section{Optimized Policy Response}\label{sec:policy}
As mentioned in the introduction, an important use of sequential inference is for policy response. In this section we briefly investigate such biosurveillance decision-making using the developed inference tools.

Public health policy makers act sequentially as outbreak data is collected.
Their aim is to balance total costs which consist of morbidity costs $C_M$ associated with the pathogen, and costs $C_A$ associated with policy actions. Denote by $\phi_t \in \R$ the policy implemented at date $t$, where we encode the action space as a subset of the real line. We consider time-additive costs on a given horizon $[0,T]$ of the form
\begin{align}\label{eq:C-funcs}
C_M & :=  \int_0^T c(I_t)\,dt \quad\text{and}\\ \nonumber
C_A & :=  \int_0^T \phi_t \,dt + \sum_{\phi_t \neq \phi_{t-}} K(\Delta \phi_t),
\end{align}
where $c(I_t)$ are the morbidity costs expressed as instantaneous rates, and without loss of generality we take
$\phi_t$ to be the instantaneous cost of the respective action. The last term in \eqref{eq:C-funcs} with $\Delta \phi_t \equiv \phi_t - \phi_{t-}$ corresponds to potential additional \emph{start-up} costs $K(\cdot)$ associated with changing a policy.

In contrast to the original presentation, now $\phi_t$ dynamically drives the evolution of $\{I_t, M_t\}$.
The impact of policy actions can be either direct or indirect. Direct actions influence the transition rates, whereby $\Theta = \Theta(\phi_t)$. For example, quarantine can cut down infectiousness rates in a population. Indirect actions influence the transition matrix $G_M=G_M(\phi_t)$ of $\{M_t\}$, e.g.~making low-season regime $M_t =1$ more likely. These can be interpreted as prophylactic measures that mitigate the seasonal effect and contain the disease at its baseline morbidity. In either case, since applied policies must be based only on currently available information, the control $\{ \phi_t\}$ is required to be $\F^I_t$-adapted.

We assume that the overall objective is to minimize average (expected) costs over the interval $[0,T]$ across all potential dynamic policies,
\begin{align}
\inf_{(\phi_t)} \;&\E^\phi \left[ C_M + \lambda C_A  \right] 
\label{eq:control-obj} &= \inf_\phi \E^\phi \left[ \int_0^T c(I_t) + \lambda \{\phi_t  +  K( \Delta \phi_t) \} \,dt \right],
\end{align}
where $\lambda$ is a Lagrange multiplier and to emphasize the impact of $\phi$ we denote the resulting probability measure as $\PP^\phi$. As $\lambda$ is increased, the budget constraints tighten; for $\lambda=0$ the optimization is solely about minimizing expected morbidity.

The optimization problem \eqref{eq:control-obj} is a partially observed stochastic control problem. The presence of unobserved quantities makes it nonstandard; a general approach is to reduce it to a standard setting by passing from $I_t$ to the augmented hyperstate $\cX_t$. As we have seen before, $\{\cX_t\}$ is finite dimensional if at least one of $\{M_t\}$ or $\Theta$ is observed, but is infinite-dimensional in the main case of interest (c) requiring joint inference. Using the Markov structure, the optimal policy response $\phi^*_t$ at time $t$ is a function of the posterior state $\cX_t$, $\phi^*_t = \Phi(I_t, \cX_t,\phi_{t-})$, for some \emph{strategy rule} $\Phi$.  Analytic treatment of such infinite-dimensional control problems is generally intractable; see \cite{LN10,LudkovskiRobustPoisson} for flexible numerical approximations that also rely on SMC.

Rather than carry out a full optimization, we investigate below some simple heuristics for such rules $\Phi$.
As our case study we consider a simple example with a binary policy response, $\phi_t \in \{0,1\}$ with $\phi_t=1$ indicating the implementation of preventive measures at date $t$. We assume that transition rates $\Theta$ are fixed but policy makers can influence the environment $\{M_t\}$ through
$$
G_M(t) \Big|_{\phi_t = 0} \! = \begin{pmatrix}
 - 6 & 6 \\ 2 & -2
\end{pmatrix} \quad \text{and }\quad G_M(t) \Big|_{\phi_t = 1} \!= \begin{pmatrix}
 -1 & 1 \\ 8 & -8
\end{pmatrix} .
$$
Thus, when counter-measures are enacted, $\{M_t\}$ is much more likely to be in the low state. Indeed, while without any actions the high season is expected to last 6 months, with mitigation it will only last an average of $52/8 \simeq 6$ weeks.

With an on/off decision-making, the dynamic policy rule is described through its action regions $D_0$ and $D_1$, such that a response is initiated as soon as $(I_t,\cX_t) \in D_1$ and $\phi_{t-}=0$, and terminated as soon as $(I_t,\cX_t) \in D_0$ an $\phi_{t-}=0$. Note that we expect to have a hysteresis region where the existing policy, whatever it may be, is continued, since there is no sufficiently strong evidence to change the response. To compare, we consider three potential classes of policies that are distinguished by the information used:
\begin{description}
\item[The infecteds-based control] relies directly on $I_t$. Since start (resp.~end) of outbreaks is characterized by persistent upward (resp.~downward) trend, we take
 $D^{Infd}_1 = \{ I_t - \mathcal{I}^{(\kappa)}_t > \bar{I}\}, D^{Infd}_0 = \{ I_t - \mathcal{I}^{(\kappa)}_t < \underline{I}\}$, where $$\mathcal{I}^{(\kappa)}_t := \frac{1}{t} \int_0^t e^{-\kappa s}I_s \,ds $$ is an exponentially weighted moving average of the number of infecteds using discount weight $\kappa$. Thus, measures are initiated when $I_t$ is sufficiently above its moving average (strong upward trend), and stopped when $I_t - \mathcal{I}^{(\kappa)}_t$ is sufficiently negative. From our experiments, using a 14-day moving average offers a good way of capturing trends.
This is a simple policy that requires no inference and can be seen as rule of thumb to provide response when $\{I_t\}$ is growing.

\item[The Bayesian policy] incorporates the full history of observations and the prior beliefs through the posterior distribution $\cX_t$. As a simple choice we consider only the posterior of $\{M_t\}$ and analyze $D^{Bayes}_1 = \{ \Pi^2_t \ge \bar{\pi} \}$ and $D^{Bayes}_0 = \{ \Pi^2_t \le \underline{\pi}\}$. Thus, countermeasures are started once the posterior probability of $\{M_t=2\}$ is above $\bar{\pi}$ (overwhelming evidence of a high season), and are stopped once the posterior probability is below $\underline{\pi}$.

\item[The Oracle policy] takes $D^{Orcl}_0 = \{ M_t = 1\}$, $D^{Orcl}_1 = \{ M_t = 2\}$, or $\phi_t = \mathbbm{1}_{\{ M_t = 2\}}$, i.e.~the control is applied precisely during the high season.  This is an idealized benchmark since it assumes that the policy-maker actually observes the latent seasonality variable. We note that it is still not the globally optimal benchmark since in principle it may not be worthwhile to respond \emph{immediately} as soon as the high season begins (e.g.~if $I_t$ is still low then ostensibly $M_t$ could revert back to its low state quickly without the need for costly interventions).

\end{description}
Clearly, the above list is not exhaustive and is for illustration purposes only. An infinite variety of further rules can be constructed. For example, including posterior information about $\Theta$ is clearly relevant to understand the severity of the outbreak and its likely future course. In general, an automated way to build an optimal policy is more appropriate than heuristics, albeit at a cost of losing some of the simplicity and intuition for the decision maker.

Once a class of policies is chosen, the decision rule needs to be optimized by minimizing over the described parametric forms such as $(\underline{\pi}, \bar{\pi})$ or $(\underline{I},\bar{I})$. Note that even for a fixed rule, the expected costs are not analytically available since the distribution of $I_t$ or $M_t$ under $\mathbb{P}^\phi$ is not explicitly computable.
We therefore resort to predictive analysis via Monte Carlo. Namely, fixing the policy rule, we simulate a large number (several hundred in our example below) of scenarios, i.e.~trajectories of $(I_t,M_t,\phi_t)$, and then average the resulting scenario costs to approximate $\E^{\phi}[ C_M + \lambda C_A]$.

Simulation under the controlled measure $\PP^{\phi}$ is straightforward thanks to the strong Markov properties of $\{\cX_t\}$ and is summarized in Algorithm \ref{algo:SIS-control}. For simplicity we restrict to the case where $\phi_t$ only changes at event times $\tau_k$. For these simulations we draw the actual parameters $\Theta$ independently from the given prior $p(\Theta)$, i.e.~assuming the model is correctly specified and then generate $\{ I_t, M_t\}$ using the Gillespie algorithm.

\begin{algorithm}[h]
\caption{Simulation of a controlled epidemic model \label{algo:SIS-control}}
{\fontsize{10.5}{13}
\begin{algorithmic}[1]
\REQUIRE $(M_0,I_0,\cX_0,\phi_0)$
\STATE Sample outbreak parameters $\Theta \sim \cX_0$
\STATE $s \leftarrow 0$
\LOOP
\STATE Simulate the next regime change date $\sigma \ge s$ of $\{M_t\}$ using the generator $G_M(\phi_s)$
\STATE Simulate the next transition $\tau \ge s$ of $\{I_t\}$ conditional on $(M_s,\phi_s,\Theta)$
\STATE $\rho \leftarrow \sigma \wedge \tau$
\STATE Save $M_{[s,\rho]}$ and $I_{[s,\rho]}$
\STATE Using PL Algorithm \ref{algo:PL} update the filter $\{\cX_t\}$ on $[s,\rho]$
\STATE Update the policy $\phi_{\rho} \leftarrow \Phi(I_{\rho},\cX_{\rho},\phi_s)$
\STATE Set $s \leftarrow \rho$
\ENDLOOP
\end{algorithmic}}
\end{algorithm}

\subsection{Cost Functionals}\label{sec:costs}
There are many possible summary statistics to evaluate the relative merit of different mitigation strategies. Among morbidity measures, one can consider average number of infecteds $\E^\phi[ \frac{1}{T} \int_0^T I_t \,dt]$, maximum infecteds $\E^\phi[ \max_{0 \le t \le T} I_t ]$, or the proportion of time that $I_t$ is above some level $I_{high}$, $\E^{\phi} [ \int_0^T \mathbbm{1}_{\{ I_t \ge I_{high}\}} \,dt ]$. One can also include metrics regarding the response, such as the average length of time countermeasures are enacted $\E^{\phi} [ \int_0^T \mathbbm{1}_{\{ \phi_t = 1\}} \,dt ]$, the number of times counter-measures are started $\E^{\phi} [ \sum_{t \le T} \mathbbm{1}_{\{\Delta \phi_t \neq 0\}} ]$, etc.
To illustrate, we consider the following two examples of cost functionals:
\begin{subequations}
\begin{align}
c_1(I,\phi) &:= \int_0^T I_t + 0.02 (I_t - 200)_+^2 + 50 \cdot \mathbbm{1}_{\{\phi_t = 1\}} \,dt;  \label{eq:c1} \\ 
c_2(I,\phi) &:= \int_0^T \bigl( I_t + 1000 \cdot \mathbbm{1}_{\{I_t > 300\}} + 200 \cdot \mathbbm{1}_{\{\phi_t = 1\}} \bigr)\,dt 
+ 1400 \cdot \sum_{t \le T} \mathbbm{1}_{\{\Delta \phi_t = 1\}}. \label{eq:c2}
\end{align}
\end{subequations}
The cost functional $c_1$ has a piecewise-quadratic cost in terms of the number of infecteds $I_t$. Here a ``soft'' threshold of $I_{high}=200$ infecteds is applied, as well as a basic morbidity cost that is proportional to $I_t$. This $c_1$ also mildly penalizes the amount of time counter-measures are applied through the third term in \eqref{eq:c1}. The cost functional $c_2$ is geared more towards minimizing mitigation resources. It has much higher policy costs of 200 per day and also rewards policy stability (i.e.~avoiding too many changes in policy or ``chattering''). The latter is taken into account in \eqref{eq:c2} through the penalty $K(\Delta \phi_t) = 1400 \cdot\mathbbm{1}_{\{\Delta \phi_t = 1\}}$ which imposes a start-up cost of $1400$ (week's worth of policy costs) each time the counter-measures are begun. In terms of outbreak costs, $c_2$ considers total number of infecteds-days and additionally imposes  a discontinuous penalty whenever $I_t$ exceeds 300 (i.e.~$3\%$ of the population) which can be thought of as a ``hard'' (but high) target regarding tolerable number of infecteds.

\begin{figure*}[ht]
\begin{center}
\includegraphics[height=3in]{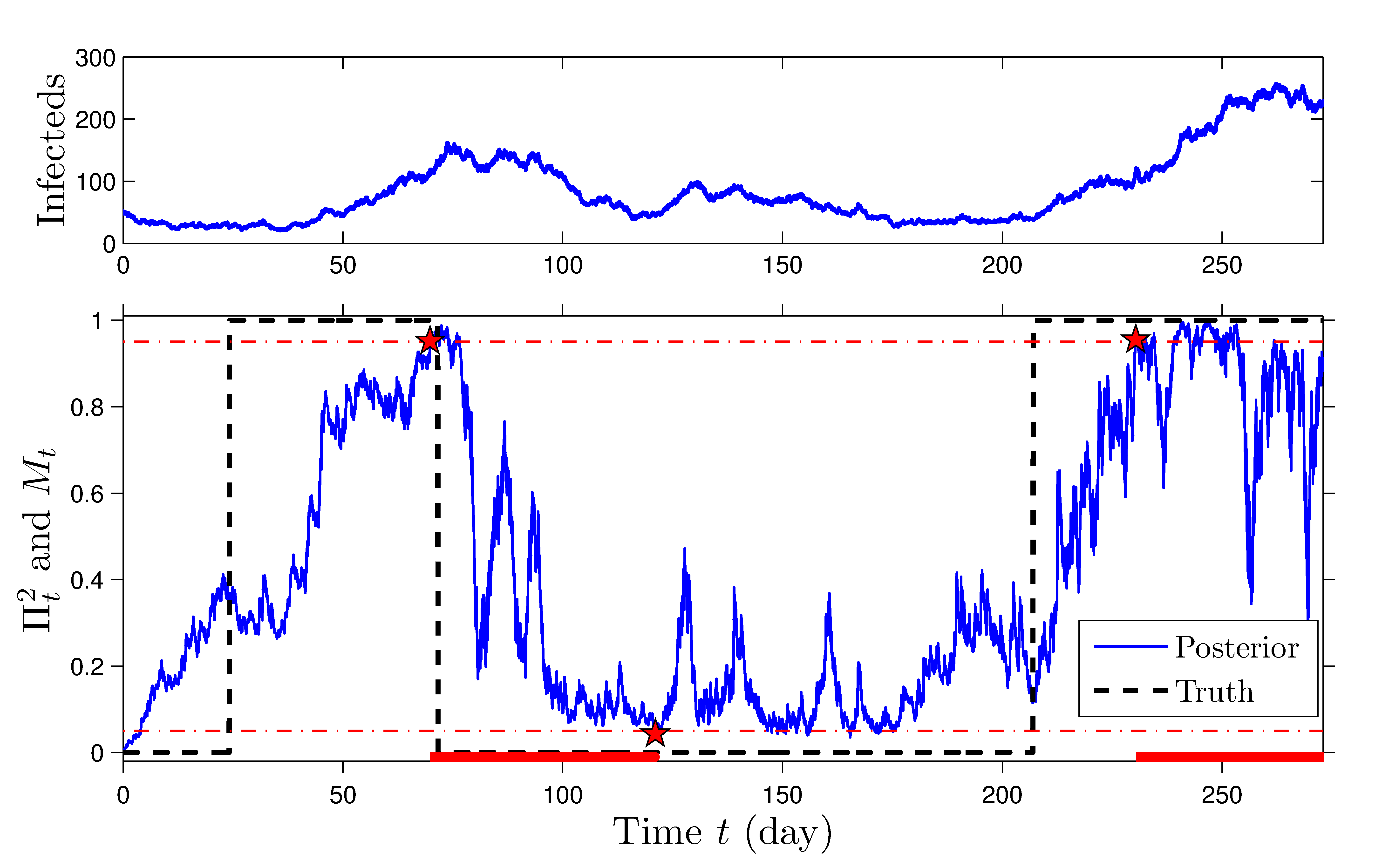}
\caption{Dynamic Bayesian control of $\{M_t\}$ over a sample trajectory of the outbreak. The top panel shows the infecteds numbers $\{I_t\}$; the bottom panel shows the true seasonal factor $\{M_t\}$ and the posterior probability of high season $\{\Pi^2_t\}$. Countermeasures are applied as soon as $\Pi^2_t > 0.95 = \bar{\pi}$ and stopped once $\Pi^2_t < 0.05 = \underline{\pi}$. The intervals of action  are indicated by the solid bars on the $x$-axis, and the times of policy changes are marked with stars. The PL algorithm used $J=5000$ particles. All other parameters are from Table \ref{tbl:params}.
}\label{fig:sample-control}
\end{center}

\end{figure*}

To generate outbreaks, we fix  $\theta_2 = 0.25$ and sample $\theta_1 \sim Ga(1700,20 \cdot 365)$, which roughly means $\theta_1 \in [0.21, 0.25]$. The remaining parameters, including $SF=0.15$, are from Table \ref{tbl:params}. Figure \ref{fig:sample-control} shows the resulting Bayesian-type dynamic policy on a sample trajectory of $\{I_t\}$. In this example, counter-measures start once the probability of being in the high season is at least 95\% and are ended once $\Pi^2_t < 0.05$ and $I_t < I_{high}=200$; the latter is to make sure that the outbreak is fully contained.  In the figure, the filter reacts rather slowly to the first outbreak, perhaps due to lower than normal numbers of infecteds at its onset. Coincidentally, once counter-measures are finally started, $\{M_t\}$ reverts back to low season almost immediately. The second outbreak begins around seven months and is responded to within $~20$ days. At the end of the simulation, $M_t=2$ remains in the high season. Note that the unrealistic assumption (solely for simplicity of presentation) that $\{M_t\}$ goes back to its original transition matrix once $\phi_t=0$ (even after counter-measures were applied previously) implies that multiple high seasons are likely to occur over the nine months.

\begin{table*} [ht]
\begin{center}
\begin{tabular}{|@{}lr|rrrr|rr|}\hline 
 &  & Ave.& Policy  & Ave.~Days  & Ave.~Policy & $\E^{\phi} [ c_1]$ &  $\E^{\phi} [ c_2]$ \\
$\,$ Policy       &            & $I_t$ & Freq. & $I_t > 300$  & Start-Ups & ('000s)  & ('000s)    \\
$\,$ Type & Thresholds  & $(\pm 3)$ & $(\pm 1\%)$ & $(\pm 1.5)$ & $(\pm 0.1)$ & $(\pm 4)$& $(\pm 2.5)$ \\
 \hline \hline
$\,$ Baseline & (Do-Nothing) & 174 & 0\% & 54  & 0  & 187 & 102\\
$\,$ Oracle 	& & $95$	& 41\% & 15  & 4.8 & 60 & 70 \\ \hline
$\,$ Infecteds	&	(20,-10) & 92 & 37\% & 13 & 2.6& 50 & 66 \\
$\,$ Infecteds	&	(40,-20) & 106 & 28\% & 17 & 5.3 & 59  & 65 \\ \hline
$\,$ Bayesian & (0.80,0.01) & 91 & 49\% & 12 & 2.7 & 48 & 67 \\
$\,$ Bayesian & (0.95,0.05) & 108 & 30\% & 19 & 2.6 & 56 & 64  \\
\hline  
\end{tabular}
\caption{\label{tbl: costs} Expected costs and summary statistics of selected response strategies over 500 simulated scenarios. The chosen Infecteds-based rule is $D^{Infd}_1 = \{ I_t - \mathcal{I}^{(14)}_t > \bar{I} \}, D^{Infd}_0 = \{ I_t - \mathcal{I}^{(\kappa)}_t < \underline{I} \cap I_t < I_{high} \}$ and the chosen Bayesian rule is $D^{Bayes}_1 = \{ \Pi^2_t \ge \bar{\pi} \}$ and $D^{Bayes}_0 = \{ \Pi^2_t \le \underline{\pi} \cap I_t < I_{high} \}$ for the specified thresholds. Bayesian algorithms used $J=3000$ particles. The Baseline policy is to do nothing $\phi_t \equiv 0$, and the Oracle policy is $\phi_t = \mathbbm{1}_{\{ M_t = 2\}}$. Standard deviations of the computed expected values are in brackets.}
\end{center}
\end{table*}

The performance of several mitigation policies is compared in Table \ref{tbl: costs}. The table demonstrates that depending on the priorities of the policy makers, different mitigation strategies should be considered. Crucially, one must consider the trade-off between fast response and potentially unnecessary interventions.  This trade-off is clearly observed with some strategies being more aggressive (and hence producing lower expected morbidity) and other strategies being more conservative. For instance, the Bayesian policy with $(\underline{\pi},\bar{\pi}) = (0.01,0.8)$ is much more aggressive than a similar policy with $(\underline{\pi},\bar{\pi}) = (0.05,0.95)$, as it starts counter-measures as soon as there is at least 80\% chance of being in the high season, and continues them until the posterior probability drops below 1\% (compared to starting at $\Pi^2_t > 95\%$ and stopping as soon as $\Pi^2_t < 5\%)$. Not surprisingly, it reduces average number of infecteds by nearly 15\% and the  average number of days when there are more than 300 infecteds by over 30\% in comparison. This comes at a cost of over 40 additional days on average when countermeasures are applied. Which policy is better therefore depends on the weightings (cf.~Lagrange multiplier $\lambda$ in \eqref{eq:control-obj}) placed on the different ingredients of the cost functional; here $\E^\phi[c_1]$ is smaller for
$(\underline{\pi},\bar{\pi}) = (0.01,0.8)$, while $\E^\phi[c_2]$ is smaller for $(\underline{\pi},\bar{\pi}) = (0.05,0.95)$.

Overall, we find that strategies based on Bayesian inference seem able to outperform policies based on $I_t$ only, though the difference is not very statistically significant. Of course, this result must be tempered as we did not perform an exhaustive optimization and it obviously depends on the cost functionals considered.
 We also note that strategies differ a lot in achieving similar results. For instance the Bayesian policy with $(\underline{\pi},\bar{\pi}) = (0.01,0.8)$ has similar expected costs using the functional $c_1$ as the Infecteds-based policy with $(\underline{I},\bar{I}) = (-10,20)$. However, the latter is much more variable (over 5.3 expected policy start-ups) in time, though lasting shorter periods (20\% less frequently).

 As a final comparison, Table \ref{tbl: costs} points out that no action at all is clearly sub-optimal and whatever the aims of the policy makers, \emph{some} mitigation is obviously beneficial. Moreover, the idealized Oracle control is in fact not optimal either, since it does not account for the dynamics of $\{I_t\}$ and by tracking $\{M_t\}$ exactly tends to act/end too quickly.
Further tailoring and refinement of risk metrics is obviously needed for practical use and raises a host of interesting inter-disciplinary questions that will be explored in upcoming works.

\bibliographystyle{spbasic}
\bibliography{hmskmLudkovskiLin}

\begin{thebibliography}{42}
\providecommand{\natexlab}[1]{#1}
\providecommand{\url}[1]{{#1}}
\providecommand{\urlprefix}{URL }
\expandafter\ifx\csname urlstyle\endcsname\relax
  \providecommand{\doi}[1]{DOI~\discretionary{}{}{}#1}\else
  \providecommand{\doi}{DOI~\discretionary{}{}{}\begingroup
  \urlstyle{rm}\Url}\fi
\providecommand{\eprint}[2][]{\url{#2}}

\bibitem[{Amrein and K{\"u}nsch(2012)}]{AmreinKunsch12} 
Amrein M, K{\"u}nsch H (2012) 
Rate estimation in partially observed {M}arkov jump processes with measurement errors. 
Stat. Comp. 22(2):513--526

\bibitem[{Andersson and Britton(2000)}]{AnderssonBrittonBook}
Andersson H, Britton T (2000) Stochastic epidemic models and their statistical
  analysis, Lecture Notes in Statistics, vol 151. Springer, New York


\bibitem[{Ball and Neal(2002)}]{BallNeal02}
Ball F, Neal P (2002) A general model for stochastic {SIR} epidemics with two
  levels of mixing. Math Biosci 180:73--102
  
\bibitem[{Bellomy(2011)}]{Sbc-Phd-report}
Bellomy A (2011) Influenza surveillance activities local update report. Tech.
  rep., Santa Barbara County Public Health Department,
  \url{http://www.countyofsb.org/phd/epi.aspx?id=23610&ekmensel=15074a7f_1152_1272_23610_1}

\bibitem[{Boys et~al(2008)Boys, Wilkinson, and
  Kirkwood}]{Boys:Wilk:Kirk:baye:2008}
Boys RJ, Wilkinson DJ, Kirkwood TB (2008) Bayesian inference for a discretely
  observed stochastic kinetic model. Stat. Comp. 18(2):125--135

\bibitem[{Capp{\'e} et~al(2005)Capp{\'e}, Moulines, and Ryd{\'e}n}]{RydenBook}
Capp{\'e} O, Moulines E, Ryd{\'e}n T (2005) Inference in hidden {M}arkov
  models. Springer Series in Statistics, Springer, New York


\bibitem[{Carvalho et~al(2010)Carvalho, Johannes, Lopes, and
  Polson}]{CarvalhoPL10}
Carvalho CM, Johannes M, Lopes HF, Polson N (2010) Particle learning and
  smoothing. Stat. Sci. 25:88--106

\bibitem[{Carvalho et~al(2011)Carvalho, Johannes, Lopes, and
  Polson}]{CarvalhoPL11}
Carvalho CM, Johannes M, Lopes HF, Polson N (2011) Particle learning for
  sequential {B}ayesian computation. Bayes. Stat. 9:317--360

\bibitem[{Cauchemez and Ferguson(2008)}]{CauchemezFerguson08}
Cauchemez S, Ferguson N (2008) Likelihood-based estimation of continuous-time
  epidemic models from time-series data: application to measles transmission in
  london. J. R. Soc. Interface 5(25):885--897

\bibitem[{Chowell et~al(2009)Chowell, Viboud, Wang, Bertozzi, and
  Miller}]{Chowell09adaptive}
Chowell G, Viboud C, Wang X, Bertozzi S, Miller M (2009) Adaptive vaccination
  strategies to mitigate pandemic influenza: Mexico as a case study. PLoS One
  4(12):e8164

\bibitem[{Cintron-Arias et~al(2009)Cintron-Arias, Castillo-Chavez, Bettencourt,
  Lloyd, and Banks}]{CintronArias09}
Cintron-Arias A, Castillo-Chavez C, Bettencourt L, Lloyd A, Banks H (2009) The
  estimation of the effective reproductive number from disease outbreak data.
  Math. Biosci. Eng. 6 (2):261--282

\bibitem[{Doucet et~al(2001)Doucet, de~Freitas, and Gordon}]{deFreitas}
Doucet A, de~Freitas N, Gordon N (eds)  (2001) Sequential {M}onte {C}arlo
  methods in practice. Statistics for Engineering and Information Science,
  Springer, New York

\bibitem[{Dukic et~al(2010)Dukic, Lopes, and Polson}]{DukicPolsonPL}
Dukic V, Lopes H, Polson N (2010) Tracking flu epidemic using {G}oogle {F}lu
  {T}rends and particle learning. Available at
  http://www.ssrn.com/abstract=1513705

\bibitem[{Dushoff et~al(2004)Dushoff, Plotkin, Levin, and
  Earn}]{DushoffPlotkin04}
Dushoff J, Plotkin J, Levin S, Earn D (2004) Dynamical resonance can account
  for seasonality of influenza epidemics. Proc. Nat. Acad.
  Sci. USA 101(48):16,915

\bibitem[{Gilks and Berzuini(2001)}]{Gilk:Berz:foll:2001}
Gilks WR, Berzuini C (2001) Following a moving target: {M}onte {C}arlo inference
  for dynamic {B}ayesian models. J. Roy. Stat. Soc. B
  63:127--146

\bibitem[{Golightly and Wilkinson(2011)}]{GolightlyWilkinson11}
Golightly A, Wilkinson D (2011) {B}ayesian parameter inference for stochastic
  biochemical network models using particle {M}arkov chain {M}onte {C}arlo. Interface
  Focus 1(6):807--820

\bibitem[{Golightly and Wilkinson(2006)}]{Goli:Wilk:baye:2006}
Golightly A, Wilkinson DJ (2006) Bayesian sequential inference for stochastic
  kinetic biochemical network models. J. Comp. Biol.
  13(3):838--851

\bibitem[{Gordon et~al(1993)Gordon, Salmond, and
  Smith}]{Gord:Salm:Smit:nove:1993}
Gordon NJ, Salmond DJ, Smith AFM (1993) Novel approach to
  nonlinear/non-{G}aussian {B}ayesian state estimation. IEE Proc. F140:107--113

\bibitem[{Grassly and Fraser(2006)}]{GrasslyFraser06}
Grassly N, Fraser C (2006) Seasonal infectious disease epidemiology.
 Proc. Roy. Soc. B - Biol. Sci.  273(1600):2541--2550

\bibitem[{Halloran et~al(2008)Halloran, Ferguson, Eubank, Ira M.~Longini,
  Cummings, Lewis, Xu, Fraser, Vullikanti, Germann, Wagener, Beckman, Kadau,
  Barrett, Macken, Burke, and Cooley}]{HalloranPnas08}
Halloran ME, Ferguson NM, Eubank S, Ira M~Longini J, Cummings DAT, Lewis B, Xu
  S, Fraser C, Vullikanti A, Germann TC, Wagener D, Beckman R, Kadau K, Barrett
  C, Macken CA, Burke DS, Cooley P (2008) Modeling targeted layered containment
  of an influenza pandemic in the united states. Proc. Nat. Acad. Sci. USA 105 (12):4639--4644

\bibitem[{He et~al(2010)He, Ionides, and King}]{IonidesKingMeasles10}
He D, Ionides E, King A (2010) Plug-and-play inference for disease dynamics:
  measles in large and small populations as a case study. J. Roy. Soc. Interface 7(43):271--283

\bibitem[{Jewell et~al(2009)Jewell, Kypraios, Neal, and
  Roberts}]{JewellKypraiosNealRoberts09}
Jewell C, Kypraios T, Neal P, Roberts G (2009) Bayesian analysis for emerging
  infectious diseases. Bayes. Anal. 4(3):465--496

\bibitem[{Keeling et~al(2001)Keeling, Rohani, and
  Grenfell}]{KeelingRohaniGrenfell01}
Keeling M, Rohani P, Grenfell B (2001) Seasonally forced disease dynamics
  explored as switching between attractors. Physica D
  148(3-4):317--335

\bibitem[{Kuske et~al(2007)Kuske, Gordillo, and
  Greenwood}]{KuskeGordilloGreenwood07}
Kuske R, Gordillo L, Greenwood P (2007) Sustained oscillations via coherence
  resonance in {SIR}. J. Theor. Biol. 245(3):459--469

\bibitem[{Lawson(2009)}]{lawson2009bayesian}
Lawson A (2009) Bayesian disease mapping: hierarchical modeling in spatial
  epidemiology, vol~20. Chapman \& Hall/CRC, New York

\bibitem[{LeStrat and Carrat(1999)}]{LeStratCarrat}
LeStrat Y, Carrat F (1999) Monitoring epidemiologic surveillance data using
  hidden {M}arkov models. Stat. Med. 18:3463--3478

\bibitem[{Liu and West(2001)}]{LiuWest}
Liu J, West M (2001) Combined parameter and state estimation in
  simulation-based filtering. In: Sequential {M}onte {C}arlo methods in
  practice, Stat. Eng. Inf. Sci., Springer, New York, pp 197--223

\bibitem[{Ludkovski(2012{\natexlab{a}})}]{LudkovskiCDC12}
Ludkovski M (2012{\natexlab{a}}) {B}ayesian quickest detection with
  observation-changepoint feedback. In: Proceedings of the 2012 Conference on
  Decision and Control, Maui HI Dec 9-12, 2012

\bibitem[{Ludkovski(2012{\natexlab{b}})}]{LudkovskiRobustPoisson}
Ludkovski M (2012{\natexlab{b}}) {M}onte {C}arlo methods for adaptive disorder
  problems. In: Carmona R, Moral PD, Hu P, Oudjane N (eds) Numerical Methods in
  Finance, Springer Proceedings in Mathematics, vol~12, Springer, pp 83--112

\bibitem[{Ludkovski and Niemi(2010)}]{LN10}
Ludkovski M, Niemi J (2010) Optimal dynamic policies for influenza management.
  Stat. Comm. Inf. Diseases 2(1):article 5

\bibitem[{Ludkovski and Niemi(2011)}]{LN11}
Ludkovski M, Niemi J (2011) Optimal disease outbreak decisions with noisy,
  delayed observations. Preprint

\bibitem[{Ludkovski and Sezer(2012)}]{LS07}
Ludkovski M, Sezer S (2012) Finite horizon decision timing with partially
  observable {P}oisson processes. Stoch. Models 28(2):207--247

\bibitem[{Mart\'inez-Beneito et~al(2008)Mart\'inez-Beneito, L\'opez-Qu\'ilez,
  and L\'opez-Maside}]{MartinezBeneito08}
Mart\'inez-Beneito C, L\'opez-Qu\'ilez A, L\'opez-Maside A (2008) Bayesian
  {M}arkov switching models for the early detection of influenza epidemics.
  Stat. Med. 27:4455--4468

\bibitem[{Merl et~al(2009)Merl, Johnson, Gramacy, and Mangel}]{MerlGramacy09}
Merl D, Johnson R, Gramacy B, Mangel M (2009) A statistical framework for the
  adaptive management of epidemiological interventions. PLoS ONE 4(6):e5087

\bibitem[{N{\aa}sell(2002)}]{Nasell02}
N{\aa}sell I (2002) Stochastic models of some endemic infections. Math. Biosci.
  179(1):1--19

\bibitem[{Niemi(2009)}]{NiemiThesis}
Niemi J (2009) Bayesian analysis and computational methods for dynamic
  modeling. PhD thesis, Duke Univ

\bibitem[{O'Neill(2002)}]{ONeill02}
O'Neill PD (2002) A tutorial introduction to {B}ayesian inference for
  stochastic epidemic models using {M}arkov chain {M}onte {C}arlo methods. Math.
  Biosci. 180:103--114

\bibitem[{Stone et~al(2007)Stone, Olinky, and Huppert}]{StoneOlinky07}
Stone L, Olinky R, Huppert A (2007) Seasonal dynamics of recurrent epidemics.
  Nature 446(7135):533--536

\bibitem[{Storvik(2002)}]{Stor:part:2002}
Storvik G (2002) Particle filters in state space models with the presence of
  unknown static parameters. {IEEE} Tran. Signal Proces.
  50(2):281--289

\bibitem[{Tanner et~al(2008)Tanner, Sattenspiel, and
  Ntaimo}]{TannerSattenspielNtaimo08}
Tanner MW, Sattenspiel L, Ntaimo L (2008) Finding optimal vaccination
  strategies under parameter uncertainty using stochastic programming. Math.
  Biosci. 215(2):144--151

\bibitem[{Wilkinson(2006)}]{Wilk:stoc:2006}
Wilkinson DJ (2006) Stochastic Modelling for Systems Biology. Chapman \&
  Hall/CRC, London

\end{thebibliography}

\end{document}